\documentclass[twocolumn, balance]{aastex631}

\usepackage{graphicx}
\usepackage{amssymb}
\usepackage{natbib}
\usepackage{xspace}

\newcommand{\oiii}{[O{\scshape iii}]\xspace}
\newcommand{\oiiil}{[O{\scshape iii}]$\lambda5007$\xspace}
\newcommand{\nii}{[N{\scshape ii}]\xspace}

\newcommand{\sii}{[S{\scshape ii}]\xspace}
\newcommand{\oi}{[O{\scshape i}]\xspace}
\newcommand{\neii}{[Ne{\scshape ii}]\xspace}
\newcommand{\neiii}{[Ne{\scshape iii}]\xspace}

\newcommand{\halpha}{H$\alpha$\xspace}
\newcommand{\hbeta}{H$\beta$\xspace}

\newcommand{\hda}{H$\delta_{\rm A}$\xspace}

\newcommand{\Msun}{M$_\odot$\xspace}

\newcommand{\lion}{$L_{\rm ion}$\xspace}

\begin{document}
\graphicspath{{./}{figures/}}

\title{Fading AGN in Post-Starburst Galaxies}

\author[0000-0002-4235-7337]{K. Decker French}
\affiliation{Department of Astronomy, University of Illinois, 1002 W. Green St., Urbana, IL 61801, USA} 
\author[0000-0003-1714-7415]{Nicholas Earl}
\affiliation{Department of Astronomy, University of Illinois, 1002 W. Green St., Urbana, IL 61801, USA} 
\author[0000-0003-3097-5207]{Annemarie B. Novack}
\affiliation{Department of Astronomy, University of Illinois, 1002 W. Green St., Urbana, IL 61801, USA} 
\author[0000-0002-7496-9768]{Bhavya Pardasani}
\affiliation{Department of Astronomy, University of Illinois, 1002 W. Green St., Urbana, IL 61801, USA} 
\author[0009-0003-6250-1396]{Vismaya R. Pillai}
\affiliation{Department of Astronomy, University of Illinois, 1002 W. Green St., Urbana, IL 61801, USA} 
\affiliation{Department of Physics, The University of Hong Kong, Pokfulam Road, Hong Kong}
\author[0000-0002-6582-4946]{Akshat Tripathi}
\affiliation{Department of Astronomy, University of Illinois, 1002 W. Green St., Urbana, IL 61801, USA} 
\author[0000-0003-1535-4277]{Margaret E. Verrico}
\affiliation{Department of Astronomy, University of Illinois, 1002 W. Green St., Urbana, IL 61801, USA}

\begin{abstract}
The role of AGN in quenching galaxies and driving the evolution from star-forming to quiescent remains a key question in galaxy evolution. We present evidence from the Mapping Nearby Galaxies at APO (MaNGA) survey for fading AGN activity in 6/93 post-starburst galaxies. These six galaxies show extended emission line regions (EELRs) consistent with ionization from past AGN activity, analogous to ``Hanny's voorwerp" and other systems where the [O{\scshape iii}]$\lambda5007$ emission is bright enough to be visible in broadband imaging. Using the infrared luminosities from \textit{IRAS} to estimate the current AGN luminosity, we find that 5/6 of the post-starburst galaxies have current AGN which have faded from the peak luminosity required to have ionized the EELRs. Given the rate at which we observe EELRs, the typical EELR visibility timescale, and an estimate of how often EELRs would be visible, we estimate the duty cycle of AGN activity during the post-starburst phase. The timescale for the galaxy to cycle between peaks in AGN luminosity is $t_{\rm EELR}\sim1.1-2.3\times10^5$ yr. Given the rate at which we observe current AGN activity during this phase, we estimate that the AGN spends only 5.3\% of this time (or $t_{\rm ON} = 0.6-1.3\times10^4$ yr) in its luminous phase, with the rest of the time spent ``off" or in a low-luminosity phase. The length of this duty cycle may explain why so few luminous AGN have been observed during the post-starburst phase, despite evidence for AGN feedback at work. 

\end{abstract}

\section{Introduction}

Observations of quiescent galaxies, their growth through cosmic time, and their early appearance in the Universe indicate that they must eventually stop growing via star formation. Galaxies exhibit well-known bimodal distributions in color, morphology, and gas content, dividing into blue, disky, gas-rich star-forming galaxies and red, elliptical, gas-poor quiescent galaxies \citep{Strateva2001, Baldry2004, Faber2007,Wuyts2011, Saintonge2011}. 
While some galaxies can transition from star-forming to quiescent slowly over many billions of years, others evolve rapidly over timescales less than a Gyr. At redshift $z\sim2$, one in three quiescent galaxies has experienced a recent end or ``quench" in its star formation history \citep{Belli2019}. By redshift $z\sim3$, more than half of quiescent galaxies have been rapidly quenched \citep{DEugenio2020}. These galaxies match in mass, size, and large-scale clustering with starbursting galaxies bright in the sub-millimeter, consistent with their ``post-starburst" star formation histories \citep{Toft2014, Wild2020, Wilkinson2021}.

Feedback from Active Galactic Nuclei (AGN) can in theory provide enough energy to remove the molecular gas that fuels star-formation, effectively ending growth from star formation in quiescent galaxies. 
The accretion energy available from AGN is more than enough to unbind the gas in a galaxy and end star formation \citep{Silk1998}, and the addition of AGN feedback to simulations can produce quiescent galaxies \citep{Springel2005}, but the effectiveness of this process depends heavily on how the energy is coupled to the gas \citep{Hopkins2012b}. The high star formation rates observed in many AGN host galaxies \citep[e.g.,][]{Florez2020} indicates that the effect of AGN on star formation is not simple and not instantaneous. Due to the angular momentum loss required for the gas fueling star formation to reach the nucleus, the peak in AGN activity may be several hundred Myr to more than a Gyr after the peak in starburst activity, comparable to the timescale over which a galaxy will be identifiable as post-starburst \citep{Davies2007, Schawinski2009, Wild2010, Hopkins2012b, Cales2015}. Thus, in order to study the impact of AGN on ending star formation in galaxies, we must look to galaxies where star formation has recently ended.

Multi-wavelength studies of high redshift $z>2$ post-starburst galaxies are limited due to their faint observed fluxes; spectroscopic studies have only been possible for a few dozen objects \citep[e.g.,][]{Wild2020}. However, large spectroscopic surveys at low redshift have enabled the discovery of thousands of low redshift analog post-starburst galaxies \citep{Zabludoff1996, Goto2005, Wild2007,Wild2009,Alatalo2016a}. Both low and high redshift post-starburst galaxies have had large fractions of their stellar mass formed in the last $\sim$ Gyr, with starbursts ending on similar timescales \citep{Wild2020}. The mechanisms that trigger the starbursts are likely to be similar between the low and high redshift samples. Most post-starburst galaxies at low redshift are in the field or in poor groups \citep{Zabludoff1996, Hogg2006}, not in the densest cluster environments, and dense clusters will be even more rare at higher redshift. The post-starburst fraction rises alongside the merger rate as redshift increases \citep{Snyder2011}, and while gas-rich galaxy compaction \citep{Zolotov2015} may act as an additional mechanism to trigger starbursts at high redshift, mergers may be required to start compaction. Low redshift post-starburst galaxies are thus useful proxies for high redshift massive quiescent galaxies, for studying the physics of quenching.

Observations of the molecular gas remaining in post-starburst galaxies have shown large molecular gas fractions, more similar to star-forming galaxies than other quiescent galaxies. Studies at both low redshift \citep{Rowlands2015,French2015, Alatalo2016b, Smercina2022} and high redshift \citep{Suess2017, Bezanson2022} show lowered star formation efficiencies (traced using the ratio of SFR to molecular gas masses or surface densities) during this phase. Lowered star formation efficiencies could either arise from excess molecular gas compared to the current SFRs or underestimated SFRs due to high levels of dust obscuration. While some post-starburst galaxies can have high obscured star formation rates inferred from their total IR luminosities \citep{Baron2022b}, mid-IR line tracers such as \neii + \neiii \citep{Ho2007} trace star formation on shorter timescales and show suppressed star formation efficiencies for post-starburst galaxies when available \citep{Smercina2022, French2023}. These observations indicate that starbursts can end without the complete removal or consumption of the molecular gas supply, but raise the question of why these galaxies are no longer forming stars. AGN feedback may act well into the post-starburst phase, supplying energy to suppress star formation and ultimately deplete the gas. Both the molecular gas and dust fractions are observed to decline during the post-starburst phase \citep{French2018,Li2019, Bezanson2022}, at rates higher than can be explained by the low residual star formation rates. Resolved observations of the molecular gas in post-starburst galaxies show large turbulent pressures and low velocity outflows \citep{Smercina2022, French2023}. The energy available from star formation is too low to explain the observed effects on the molecular gas, yet direct evidence for AGN feedback has been elusive.

Observing unambiguous AGN activity during the post-starburst phase is difficult due to  contamination from other energetic sources like shocks and post-AGB stars and the intrinsic variability of AGN. Optical emission line ratios \citep{Baldwin1981,Kewley2006} for post-starburst galaxies typically show LINER-like emission, which can indicate either low luminosity AGN activity, shocks, or evolved stars \citep{Caldwell1996,Yan2006,Yang2006,Wild2007,Brown2009, Wild2010,Mendel2013, DePropris2014, French2015,Alatalo2016a}. These narrow line ratios do not uniquely indicate AGN activity, as especially during the post-starburst phase, both shocks and evolved stars may be the dominant ionizing sources \citep{Yan2012,Rich2015,Alatalo2016a,Belfiore2016}. X-ray searches have revealed low \citep{Brown2009, Georgakakis2008, Lanz2022} to absent \citep{DePropris2014} 0.5--7 keV X-ray emission in low redshift post-starburst galaxies. Hot dust from AGN in low redshift post-starburst galaxies is also rare, with only $\sim3$\% of post-starburst galaxies displaying WISE colors consistent with AGN \citep{Meusinger2016, Smercina2018}, yet the presence of moderately obscured AGN is more difficult to assess. AGN are observed to vary on all timescales from hours to Myr \citep[summarized e.g. in ][]{Sartori2018}, complicating efforts to measure the properties of AGN in large samples of galaxies \citep{Hickox2014}. This intrinsic variability means that it is not enough to study the presence of current AGN activity to infer the connection to quenching processes; we must study the full duty cycle of how often AGN are active and for how long.

AGN have been observed to vary dramatically from quasar-like to LINER-like on timescales of 10,000--100,000 years, much more quickly than galaxies can evolve through the post-starburst phase, which lasts nearly $10^9$ years. Past AGN activity can illuminate clumps of gas tens of thousands of light years away from the nucleus of the galaxy. Spatially-resolved spectroscopic observations of AGN host galaxies have shown Extended Emission Line Regions (EELRs; also known as Extended Narrow Line Regions (ENLRs)) \citep[e.g.,][]{Greene2011, Liu2013, Harrison2014, Chen2019ENLR} on scales $\sim 10-100$ kpc, well beyond typical narrow line regions. Extreme examples of these regions have also been found in broad band imaging, when the narrow line flux is strong compared to the continuum. A dramatic example is ``Hanny's voorwerp" \citep{Lintott2009}, where \oiiil emission from an AGN light echo is bright enough to dominate the SDSS $g$ band. In order to power the observed emission in the voorwerp, the AGN must have been quasar-like, significantly more luminous than the AGN luminosity currently observed in the nucleus of the nearby host galaxy. Follow-up work has revealed more such cases of extended ionized regions (or ``voorwerpjes") \citep{Keel2012}, observed in SDSS imaging. These galaxies have past ionizing luminosities ranging from Seyfert-like to quasar-like, and many show evidence of dramatic decreases in the AGN luminosity, similar to Hanny's voorwerp. These cases have demonstrated how extended ionized regions can be used to probe the past AGN activity of galaxies, even in cases where the AGN has faded significantly.

EELRs provide the longest duration tracer of AGN activity we have available, which can provide key constraints on the nature of AGN activity during the post-starburst phase. Furthermore, extended ionized regions are best visible in galaxies with recent mergers, as the extended or tidally-stripped gas can be illuminated, which are common in post-starburst galaxies \citep[e.g.,][]{Sazonova2021, Ellison2022}. Indeed, EELRs have been observed in several post-starburst galaxies. \citet{Schweizer2013} observed an \oiii-bright EELR around the nearby post-starburst galaxy NGC7252. MUSE observations \citep{Prieto2016} of a post-starburst galaxy revealed an \oiii-bright EELR extending 23$^{\prime\prime}$ (10 kpc at $z\sim0.02$), or $3\times10^4$ lightyears away from the nucleus, indicating it was ionized by AGN activity $\sim3\times10^4$ years ago. The nearby post-starburst galaxy M51b (NGC 5195) has an EELR that has been observed by \citet{Watkins2018} and \citet{Xu2023}. 

Samples of EELRs in post-starburst galaxies have been difficult to obtain due to the need for either very bright emission (in order to be discoverable in broad-band imaging) or wide-field integral field spectroscopy. In this work, we use observations from the Mapping Nearby Galaxies at APO (MaNGA; \citealt{Bundy2015, Yan2016}) survey to search for post-starburst galaxies with EELRs in order to measure the presence of past AGN activity during this phase of galaxy evolution. We discuss our sample selection in \S2, our methods for identifying EELRs in \S3, present results on the inferred AGN history in \S4, discuss the role of AGN activity in the context of the rapid evolution of these galaxies in \S5, and conclude in \S6. When necessary, we assume a flat cosmology with $h=0.7$, $\Omega_{\Lambda}=0.7$, and $\Omega_{m}=0.3$. 

\section{Post-starburst Galaxies in MaNGA}

\begin{figure}
\begin{center}
\includegraphics[width=0.5\textwidth]{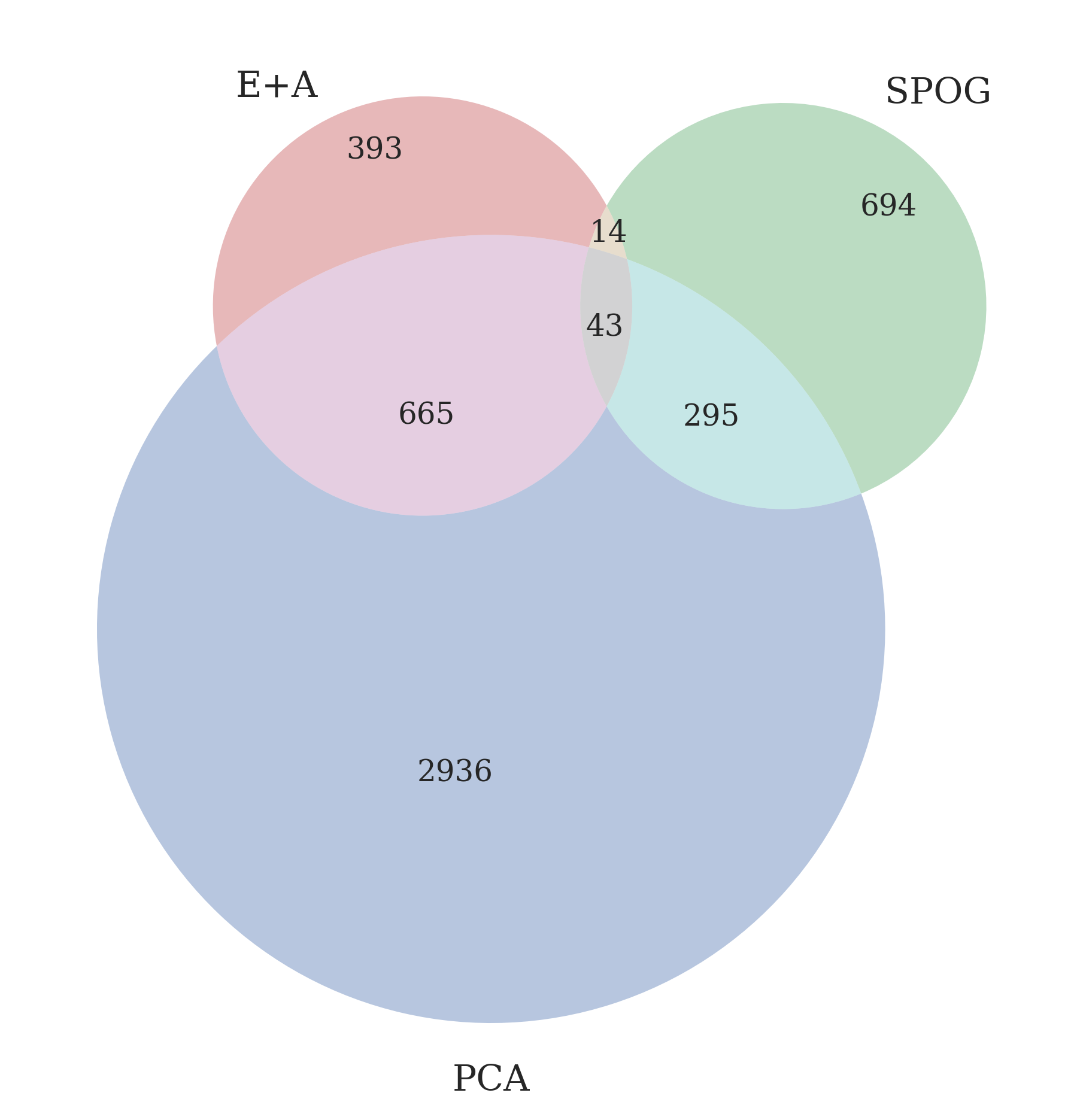}
\end{center}
\caption{Summary of the composite post-starburst galaxy sample we have constructed for this work. We select galaxies from three different samples of post-starburst galaxies selected from the SDSS in order to obtain galaxies with a variety of AGN properties and post-burst ages. We combine galaxies from the E+A sample of \citet{French2018}, the SPOG sample of \citet{Alatalo2016a}, and the PCA post-starburst sample of \citet{Wild2007,Wild2009}. Each sample overlaps with the others and our final composite sample consists of 5040 post-starburst galaxies. By combining observations across a range of samples and comparing the results of each study, we can account for the strengths and weaknesses of each selection criteria.
}
\label{fig:sample_pie}
\end{figure}

Many methods have been proposed to select galaxies evolving through the post-starburst phase \citep[see][for a recent review]{French2021}. The strengths of the Balmer absorption lines, especially H$\delta$, are used to select galaxies with a large population of A stars, indicative of a recent starburst. Selecting against current star formation is more difficult, as energy from young stars and energy from AGN activity can be difficult to fully distinguish. In this work, we combine objects from three different samples of post-starburst galaxies selected from the SDSS \citep{York2000,Strauss2002} in order to select galaxies with a variety of AGN properties and a variety of post-starburst ages.

E+A galaxies from \citet{French2018} were selected using a combination of H$\delta$ strength and a cut against H$\alpha$ emission, similar to selection methods used by \citet{Zabludoff1996, Goto2005}. This harsh cut against star formation ensures the galaxies are quiescent, and results in a low fraction ($<5$\%) of obscured starbursts \citep{Baron2022a}, but also selects against optical emission lines from AGN. The samples of Shocked POst-starburst Galaxies \citep[SPOGs][]{Alatalo2016a} and Principle Component Analysis selected galaxies \citep[PCA][]{Wild2007, Wild2009} allow for more narrow line emission. The SPOGs are selected by requiring galaxies have emission line ratios inconsistent with star formation \citep{Baldwin1981,Kewley2006}. These galaxies tend to be younger and have higher star formation rates than the E+A galaxies \citep{French2018,Ardila2018, Baron2022b}. PCA post-starbursts are selected using principle components derived from the SDSS sample, which roughly correspond to the H$\delta_{\rm A}$ and $D_n4000$ indices. Each of these samples has some overlap with the others, as shown in Figure \ref{fig:sample_pie}. We combine the objects from each of these three samples, for a total sample size of 5040. 

With this sample of 5040 post-starburst galaxies, we perform a cross-match with the MaNGA DR17 dataset, finding 93 post-starburst galaxies with MaNGA data. The MaNGA dataset contains integral field spectroscopy for over 10,000 galaxies, covering a wavelength range of 360--1000 nm with a resolution of $R\sim2000$. At a typical redshift of $z\sim0.03$, the spatial resolution corresponding to the $\approx$2.5\arcsec FWHM is $\sim1-2$ kpc. Emission line properties, kinematics, and spectral indices are available from the MaNGA Data Analysis Pipeline \citep{Westfall2019}. Our sample of post-starburst galaxies in MaNGA consists of 27 E+A galaxies, 22 SPOGs, and 89 PCA galaxies (which have significant overlap). The galaxies are found in a range of MaNGA sub-samples. There are 28 in the MaNGA primary sample, 22 in the secondary sample, and 19 in the color-enhanced sample. While MaNGA has a dedicated post-starburst galaxy ancillary program, only 20/93 of the galaxies we consider here were targeted as part of this ancillary program. Our sample of 93 contains 20/24 of the galaxies targeted as part of the post-starburst ancillary program\footnote{Though we do not include four of the post-starburst ancillary program galaxies in our analysis, we note that none of these display EELRs and their inclusion in our analysis would not substantially change our results.}. Three of the post-starburst galaxies were targeted in the ancillary WISE AGN program, and the sample contains one galaxy each from the edge on winds, pair recenter, and pair 2ifu ancillary programs.

\section{Selecting Extended Emission Line Regions}

\begin{figure*}
\begin{center}
\includegraphics[width=\textwidth]{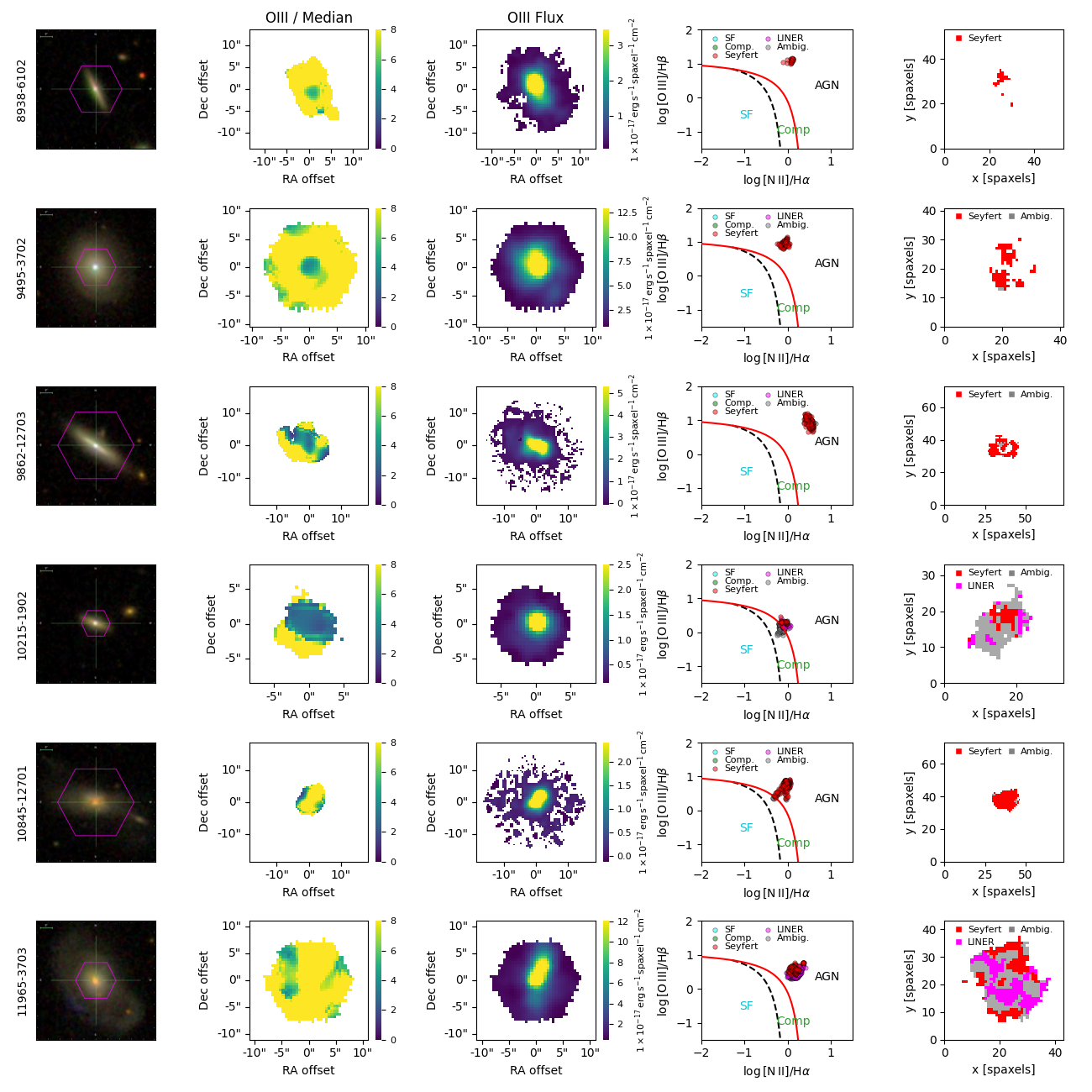}
\end{center}
\caption{In order to detect EELRs, we inspect each galaxy for regions where the ratio of \oiiil flux to the median flux is high and the emission line ratios are consistent with extended AGN ionization instead of star formation. Each row on this Figure contains the SDSS $gri$ image with MaNGA field of view overlaid, a map of the \oiiil flux to median flux, a map of the \oiiil flux, a BPT diagram, and a spatially resolved BPT classification (containing information from multiple emission line ratio classifications, see \citealt{Cherinka19}). For this last panel, spaxels with ionization indicative of Seyfert, LINER, composite, or star-formation are colored in red, magenta, green, and cyan, respectively. This Figure shows the six galaxies for which we identify EELRs. Figure \ref{fig:method2} shows six representative galaxies for which we do not identify EELRs.
}
\label{fig:method}
\end{figure*}

\begin{figure*}
\begin{center}
\includegraphics[width=\textwidth]{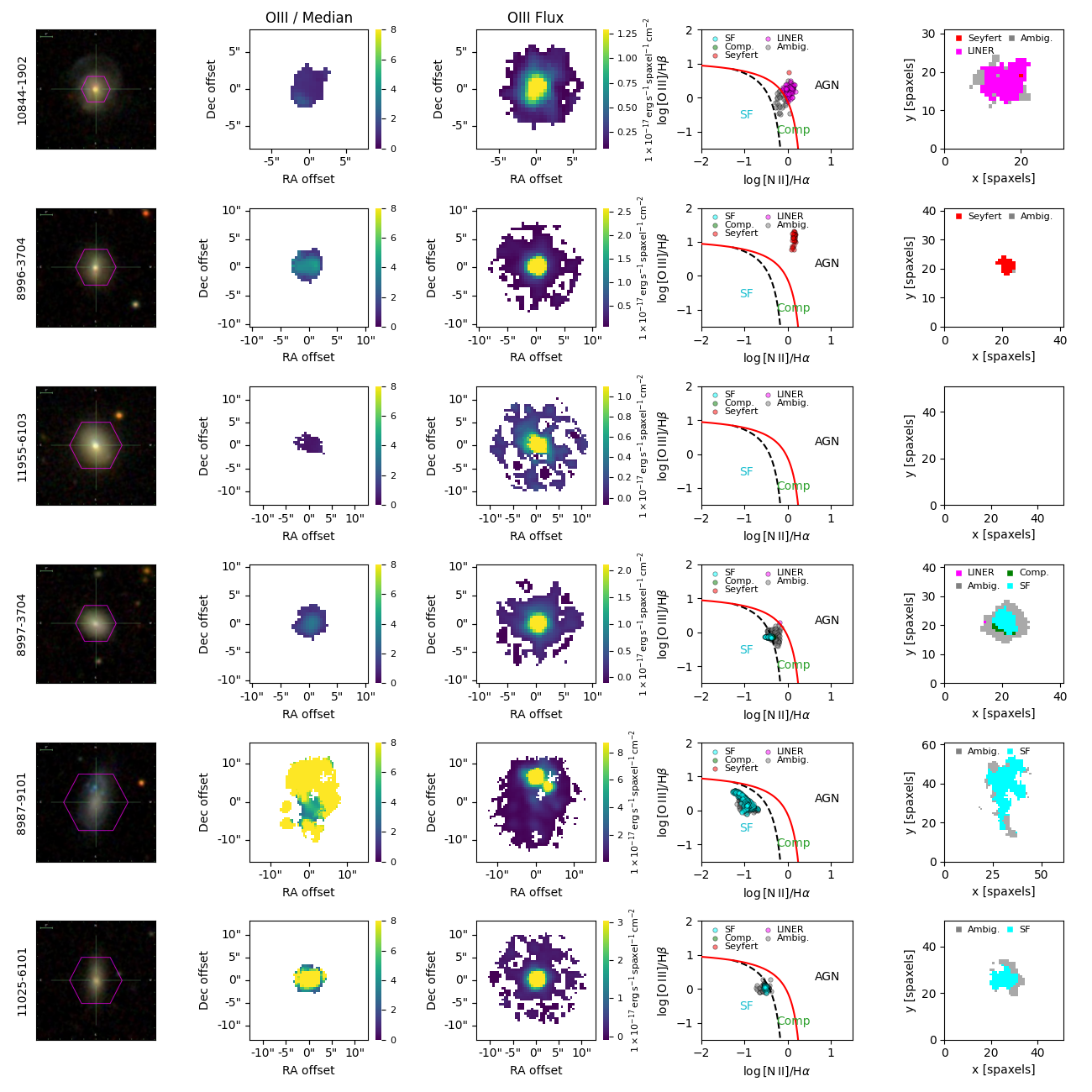}
\end{center}
\caption{Same as Figure \ref{fig:method}, but showing six representative galaxies for which we do not identify EELRs. In these cases, either we do not detect regions of high \oiiil/median flux, or the areas with high \oiiil/median are clearly dominated by ionization from star formation.
}
\label{fig:method2}
\end{figure*}

\begin{table*}[]
    \centering
    \begin{tabular}{lllllllll}
    \hline
    MaNGA PlateIFU & RA & Dec & z & E+A & SPOG & PCA & WISE AGN & MaNGA sample$^a$\\
     & (deg) & (deg) & \\
    \hline
    8938-6102 & 120.06707 & 29.471422 &  0.0453 & Y & N & N & N & Secondary\\
    9495-3702 & 123.32063 & 22.648304 & 0.0222  & Y & N & Y & N & Primary\\
    9862-12703 & 195.50057 & 27.782728 &  0.0237 & Y &  N & Y & N & Primary\\
    10215-1902 & 123.8574 & 37.3405 & 0.0397 &  N & Y & N & N & Secondary\\
    10845-12701 & 147.9806 & 3.4834 & 0.0602 & N & Y & Y & N & Secondary \\
    11965-3703 & 231.05241 & 8.544803 & 0.0371 &  N &  N & Y & Y & PSB \\
    \hline
    \end{tabular}
    \caption{Post-starburst galaxies with EELRs identified from MaNGA. $^a$ Indicates which MaNGA sample the galaxy was targeted as part of.}
    \label{tab:targets}
\end{table*}

\begin{figure*}
\begin{center}
\includegraphics[width=\textwidth]{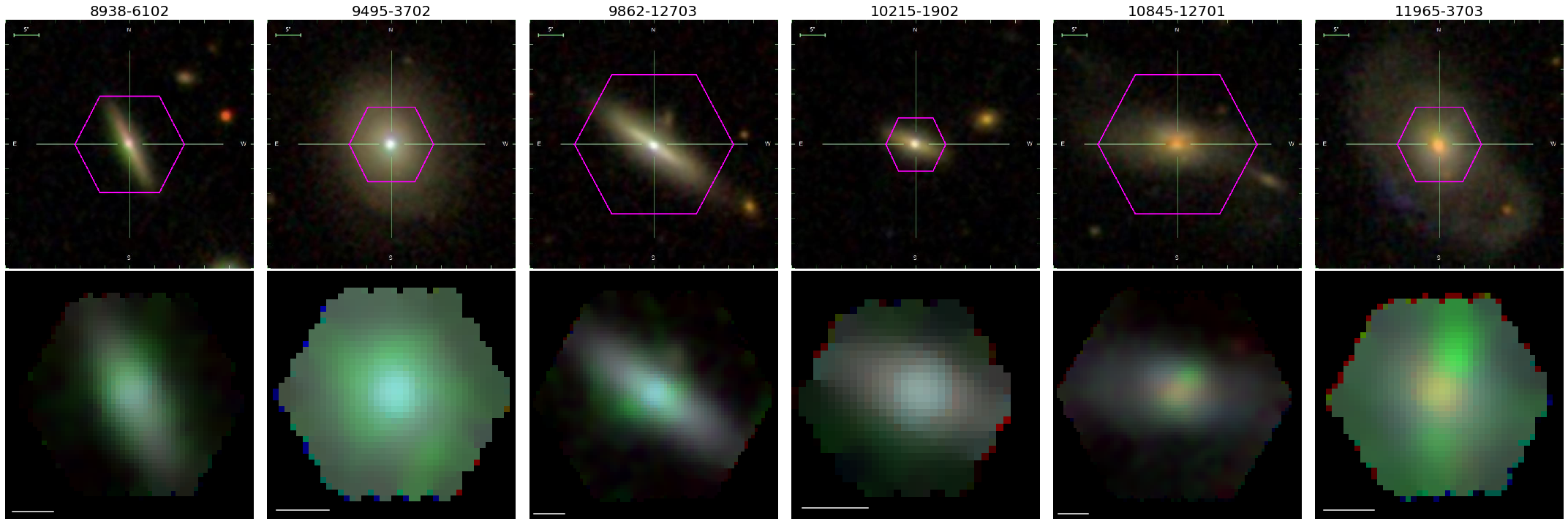}
\end{center}
\caption{EELRs in post-starburst galaxies identified using MaNGA. \textit{Top row}: SDSS $gri$ images with MaNGA fiber bundle outline overlaid in magenta. A 5 arcsec scale bar is shown in green. \textit{Bottom row}: Three color image using the average flux from MaNGA observations in two line-free continuum regions (blue: 4500-4700\AA, red: 7300-7500\AA) and the average flux in a wavelength range including the \oiiil emission line (green: 4997-5017\AA). A 5 arcsec scale bar is shown in white. These images allow for visual identification of candidate extended emission line regions, where the \oiiil line flux is strong compared to the continuum. Emission line ratio maps can then be used to exclude star forming regions. The six galaxies shown were selected from the 93 post-starburst galaxies in MaNGA by identifying regions with high ratios of \oiiil to median flux and excluding star-forming regions. These cases are analogous to Hanny's voorwerp \citet{Lintott2009} and similar systems \citet{Keel2012,Keel2017} where the ionizing flux from the nucleus on timescales $10^4-10^5$ years ago can now be seen to illuminate these extended regions. 
}
\label{fig:voorwerp_img}
\end{figure*}

In order to select galaxies with EELRs we visually inspect maps of the ratio of \oiiil flux to the median flux per spaxel. Candidate voorwerps are those with extreme ratios of \oiiil/median $\gtrsim 7$, which represents the top 5\% of \oiiil/median values for the post-starburst galaxy sample. To rule out cases where the \oiiil emission is strong due to star formation, we also inspect maps of the \oiiil/\hbeta ratio, and spatially resolved maps of emission line ratios.  We use the ratios of \nii/H$\alpha$ and \oiii/H$\beta$ \citep{veilleux87, Baldwin1981,  Kewley2001, Kauffmann2003} as well as the \sii/\halpha and \oi/\halpha ratios. Each spaxel is assigned to be Star-forming, composite, Seyfert, LINER, ambiguous, or unclassifiable \citep{Cherinka19}. We identify six post-starburst galaxies with EELRs. Properties of these galaxies are shown in Table \ref{tab:targets}. We show these maps for the six post-starburst galaxies with EELRs in Figure \ref{fig:method} and for six representative galaxies for which we do not identify EELRs in Figure \ref{fig:method2}. One of the six galaxies (11965-3703) was also identified as an EELR host in the \citet{Keel2012} sample.

In order to visualize the extended ionized regions, we construct three-color RGB images using line-free continuum regions in the red and blue, with the \oiii flux in the green channel. The six galaxies show a range of extended ionized region morphologies, with some extending out along what may be tidal tails of gas, and others extending out in biconical regions (Figure \ref{fig:voorwerp_img}). 

Of the six post-starburst galaxies with EELRs, only one (11965-3703) was targeted as a part of a MaNGA program to target post-starburst galaxies. The remaining 5/6 were part of either the MaNGA primary or secondary sample, as listed in Table \ref{tab:targets}. 

Some galaxies may have similar past AGN activity as the six post-starburst galaxies with EELRs, yet without a corresponding EELR, if they lack extended gas to be ionized. We explore this possibility in S\ref{sec:morphology}.

\section{Results}
\label{sec:calc_lion}

\subsection{Past Ionizing Luminosity}

\begin{table*}
    \centering
    \begin{tabular}{llllll}
    \hline
    MaNGA PlateIFU & log \lion$_{\rm min}$$^a$ & log \lion$_{\rm max}$$^b$ & $t_{\rm EELR}$ ($10^4$ yr)$^c$ & log $L_{\rm IR}$$^d$ & log $L_{\rm IR \ AGN}$$^e$ \\
    \hline
    8938-6102 &  $>$42.7 & $<$44.5 & 1.5 & $<43.3$ & $<43.1$\\
    9495-3702 & $>$42.9 &  $<$44.9 & 1.2 & $<42.6$ & $<42.1$ \\
    9862-12703 & $>$42.8 & $<$44.2 &  1.4 & $<42.6$ & --\\
    10215-1902 & $>$42.9 &  $<$44.5 & 1.6 & $<43.2$ & --\\
    10845-12701 & $>$43.6 &  $<$45.3 & 1.4 & $<43.6$ & --\\
    11965-3703 &  $>$43.8 &  $<$44.6 & 1.1 & 44.0 & 43.7\\
    \hline
    \end{tabular}
    \caption{EELR properties. All luminosities are in units of $\log L/($erg s$^{-1})$. $^a$ log of Peak \lion determined using \halpha, these measurements are lower limits on the true \lion. $^b$ log of upper limit on \lion determined using constraints on the density and ionization parameter. $^c$ Inferred time since peak \lion from light travel time from nucleus to peak \lion spaxel. $^d$ IR luminosity to constrain obscured current AGN luminosity. $^e$ Estimate of AGN contribution to total IR luminosity after excluding contribution from star formation (see text). }
    \label{tab:lion}
\end{table*}

We use a recombination balance assumption to estimate the past AGN activity in these galaxies. Following \citet{Keel2012, Keel2017}, we can measure a lower limit on the ionizing luminosity from the AGN by assuming the region of the galaxy traced by each spaxel is in balance with each ionization matched with a recombination series. While the \oiiil line is dominant in these regions, we use the Balmer lines \halpha and \hbeta in order to use a simple model for the radiative transfer. We correct the \halpha and \hbeta maps for dust attenuation using the Balmer decrement. By taking the count rate of H$\alpha$ or H$\beta$ photons in each spaxel, we multiply by the covering fraction of the AGN at the radius of that spaxel, and divide by the fraction of recombinations cascading through that transition, arriving at the minimum number of ionizing photons required to generate the observed count rate. 

\begin{figure*}[t!]
\begin{center}
\includegraphics[width=0.9\textwidth]{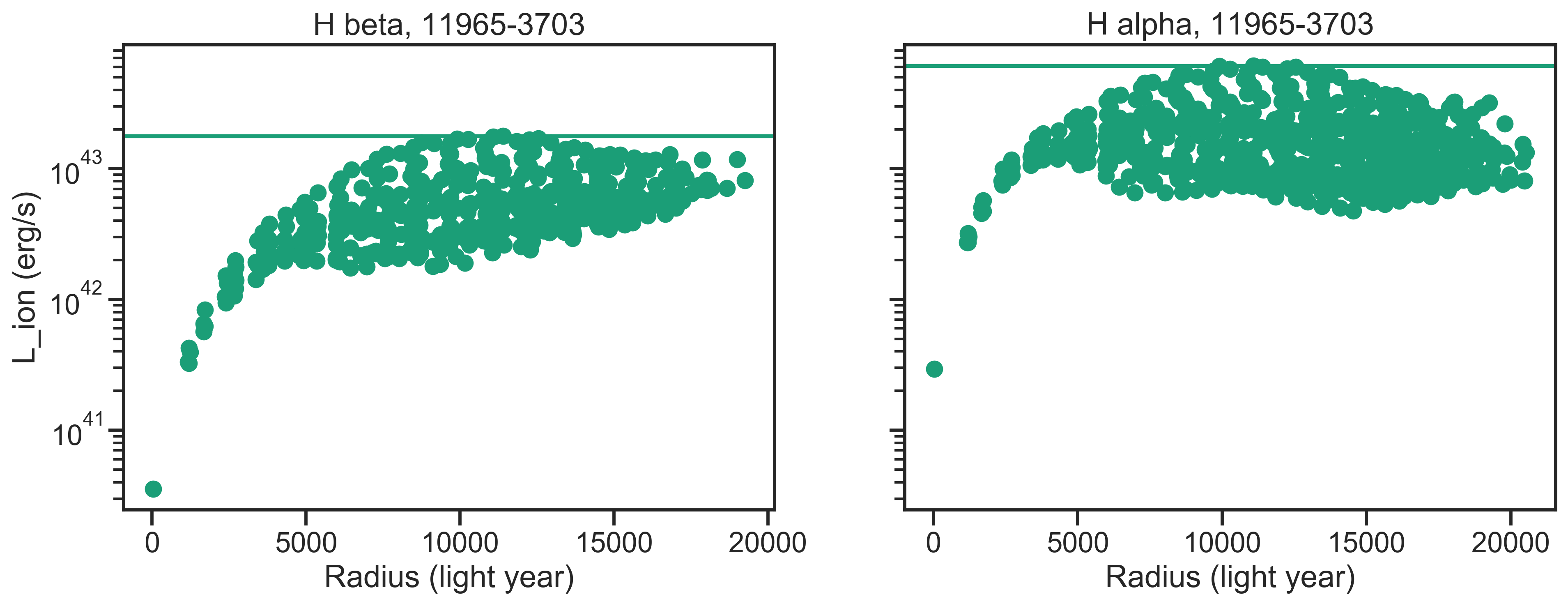}
\end{center}
\caption{Ionizing luminosity ($L_{\rm ion}$) required to provide the observed H$\alpha$ or H$\beta$ emission observed in the resolved MaNGA observations for an example post-starburst galaxy, as a function of radius. Radius is presented in units of light years to facilitate the interpretation of the AGN luminosity history (assuming the recombination time is short). Each measurement represents a lower limit to $L_{\rm ion}$, so tracing the upper envelope of each set of data provides the lower limit of $L_{\rm ion}$ as a function of radius. The maximum past $L_{\rm ion}$ is indicated with a horizontal line, and used in our analysis in Figure \ref{fig:lion_comp}.).
}
\label{fig:lion_history}
\end{figure*}

The inverse covering fraction $(1/f)$ in units of spaxels is:
\begin{equation}
    \frac{1}{f} = \frac{4 \pi}{(2\arctan(0.5/r))^2},
\end{equation}
where $r$ is the projected distance (in spaxels) from the nucleus to each spaxel. The inferred ionizing luminosity is found as:
\begin{equation}
    L_{\rm ion} = \frac{n_r L_H E_{\rm ion}}{E_H} \frac{1}{f}, 
\end{equation}
where $n_r$ is the number of ionizing photons per each recombination, with $n_r = 9.1$ for \halpha and $n_r=12.2$ for \hbeta, $L_H$ is the attenuation-corrected luminosity in each spaxel for \halpha or \hbeta, $E_{\rm ion}$ is the energy between 13.6 and 54.6 eV that we assume to be the continuum ionizing the Hydrogen\footnote{We assume photons with energy $>54.6$ eV will be absorbed primarily by Helium.}, and $E_H$ is the energy per \halpha or \hbeta photon. We show an example of the results of this measurement in Figure \ref{fig:lion_history}. 

The past inferred ionizing luminosity in the six post-starburst systems ranges from $10^{42.5}-10^{44}$ erg/s, on the low end of the AGN hosts considered by \citet{Keel2012} ($10^{42.5}-10^{45}$ erg/s). This indicates that these galaxies each hosted a Seyfert-like AGN in their recent pasts. 

To constrain the upper limit on the past ionizing luminosity in these systems, we use the [S{\scshape ii}]$\lambda6717$/[S{\scshape ii}]$\lambda6731$ ratio to estimate the electron density and the [O{\scshape ii}]$\lambda3727$/[O{\scshape iii}]$\lambda5007$ ratio to estimate the ionization potential, following \citep{Komossa1997, Bennert2005,Keel2012}. For each line ratio, we consider only the MaNGA spaxels for the EELR, selected by requiring the \oiii/median flux ratio to be $>7$. Next, we require each emission line for the ratios listed above to be detected with a SNR$>5$. We take the median [S{\scshape ii}]$\lambda6717$/[S{\scshape ii}]$\lambda6731$ flux ratio and use {\tt PyNeb} \citep{Luridiana2015} to constrain the electron density $n_e$, assuming a temperature of $T=10^4$ K. The resulting densities range from $<10$ cm$^{-3}$ to 132 cm$^{-3}$. Next, we use the [O{\scshape ii}]$\lambda3727$/[O{\scshape iii}]$\lambda5007$ ratio to estimate the ionization potential, which is independent of the ionizing continuum shape for the low density regime of these regions. The ionization potentials for these systems range from $\log U = -3.0$ to $\log U = -2.2$. These ranges of density and ionization potential are similar to those measured for the EELRs around AGN by \citet{Keel2012}. With these, we determine an upper limit on the past ionizing luminosity as,
\begin{equation}
    L_{\rm ion, \ max} = 4 \pi d^2 n_e U c E_{\rm ion}
\end{equation}
where $d$ is the distance from the nucleus to the EELR. For the six post-starburst galaxies with EELRs, we list these constraints in Table \ref{tab:lion}. The upper limits on \lion range from $10^{44.2}-10^{45.3}$. In all cases, the upper limits obtained using this method are higher than the lower limits obtained using the recombination balance method above, resulting in a physically-meaningful range of past maximum \lion. 

\subsection{Comparison of Past Ionizing Luminosity with Current Luminosity}

\begin{figure*}
\begin{center}
\includegraphics[width=0.3\textwidth]{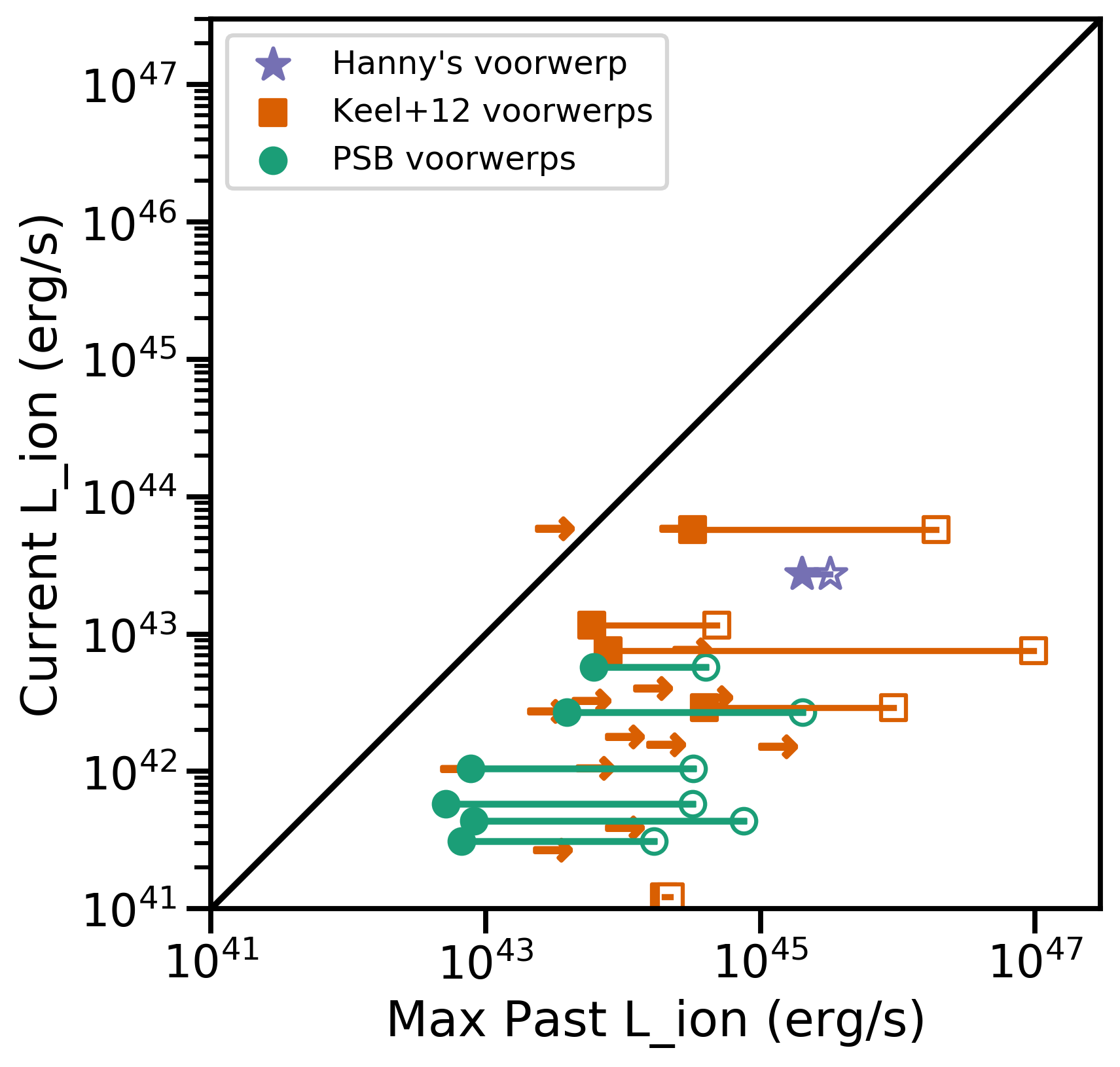}
\includegraphics[width=0.3\textwidth]{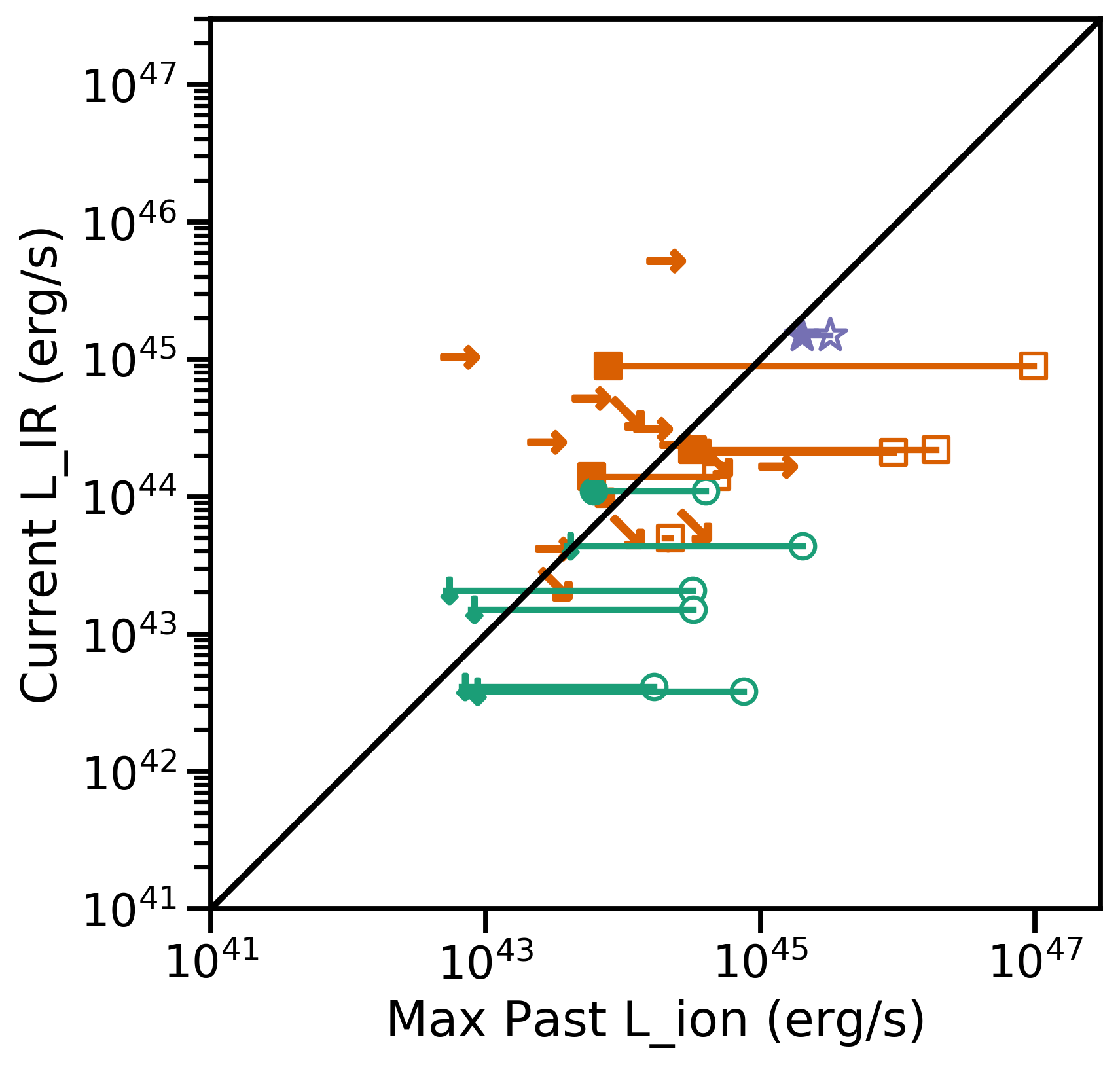}
\includegraphics[width=0.3\textwidth]{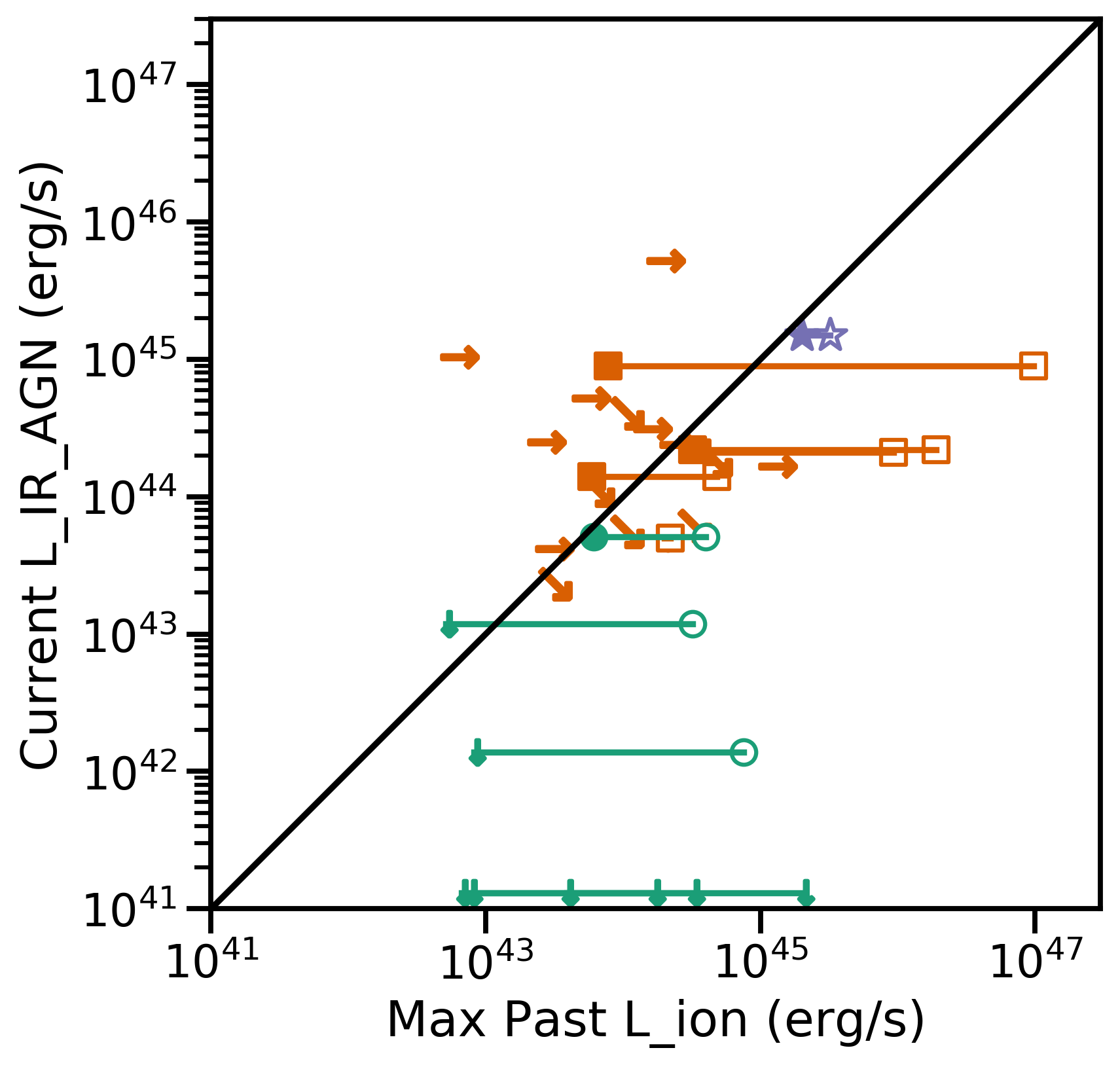}
\end{center}
\caption{\textit{Left:} Maximum past \lion vs. current \lion measured from central 3\arcsec\ SDSS fiber for post-starburst galaxies with EELRs, AGN with EELRs from \citet{Keel2012} and Hanny's Voorwerp \citep{Lintott2009}. Lower limits on the maximum past \lion obtained using recombination balance are indicated with solid plot points (obtained using \halpha for the post-starburst sample), and upper limits obtained using the ionization potential and electron density are indicated with open plot points, connected with a line. Nearly all galaxies fall below the 1:1 line, but this does not necessarily represent fading AGN activity, as the center of the galaxy may have higher dust obscuration. 
\textit{Center:} Maximum past \lion vs. current IR luminosity measured for the same samples using \textit{IRAS} 60$\mu$m and 100 $\mu$m photometry. Uncertainties on the IR luminosities are of order 10\% due to calibration uncertainties. Assuming the IR luminosity is entirely from AGN heating, galaxies below the 1:1 line are cases with definitive AGN fading. Thus, roughly half of the \citet{Keel2012} AGN and 2/6 of the post-starburst galaxies have clear signs of fading AGN activity. Next, we estimate the contribution of the IR luminosity from the star formation rates in the post-starburst galaxy sample and subtract this from the total IR luminosity to obtain the IR luminosity likely due to current AGN activity. 
\textit{Right:}  Maximum past \lion vs. current AGN IR luminosity for post-starburst galaxies (for AGN, same as center panel). 3/6 post-starburst galaxies have IR luminosities entirely consistent with the current SFRs, indicating the IR emission is not due to current AGN activity. With this correction, 5/6 post-starburst galaxies show evidence for fading AGN activity. 
}
\label{fig:lion_comp}
\end{figure*}

We compare our constraints on the past ionizing luminosity with measurements of the current ionizing luminosity in each galaxy in Figure \ref{fig:lion_comp}. First, we compare to the current \lion measured using the flux observed in the central 3\arcsec\ SDSS fiber, corrected for dust attenuation using the Balmer decrement. Nearly all galaxies fall below the 1:1 line, but this does not necessarily represent fading AGN activity, as the center of the galaxy may have higher dust obscuration, beyond what can be traced using the Balmer decrement. Following \citet{Keel2012}, we next compare to the current IR luminosity, determined using the \textit{Infrared Astronomical Satellite (IRAS; } \citealt{Neugebauer1984}) 60$\mu$m and 100 $\mu$m photometry
\begin{equation}
    f_{FIR} {\rm [W \ m^{-2}]} = 1.26\times10^{-14} \ (2.58 f_{60} +f_{100}) \ \ \\
\end{equation}
where $f_{FIR}$ is the FIR flux and $f_{60}$ and $f_{100}$ are the 60 and 100 $\mu$m fluxes in units of Jy. We use {\tt scanpi}\footnote{\url{https://irsa.ipac.caltech.edu/applications/Scanpi/}} to obtain upper limits from {\it IRAS} for the cases without significant detections, following \citet{Baron2022a}. For 5/6 of the galaxies, neither \textit{IRAS} band is detected. The IR luminosity will provide information about obscured AGN luminosity, but will also be contaminated by IR emission from star formation, so it provides an upper limit to the current \lion. Galaxies that fall below the 1:1 line in this comparison are unambiguously fading AGN. Two of the six post-starburst galaxies with EELRs, along with many of the galaxies from \citet{Keel2012} and Hanny's voorwerp fall into this category. These two post-starbust galaxies lack significant \textit{IRAS} detections, but have constraining upper limits on the IR luminosity that indicate the AGN has faded since the light ionizing the EELR was emitted. 

Finally, we attempt to correct for the contribution of star formation to the IR luminosity for the post-starburst sample. We use the SFR from the Pipe3D catalog \citep{Sanchez2018} in DR17, calculated based on the stellar population synthesis fits. We determine the IR luminosity that would be generated from this amount of inferred star formation, $L_{SF}$, as
\begin{equation}
    L_{SF}  =  10^{10} L_\odot \textrm{SFR}/(M_\odot \  {\rm yr}^{-1}), \ 
\end{equation}
assuming a Chabrier IMF following \citet{Baron2022b}. For three of the post-starburst galaxies with EELRs, the observed IR luminosity is entirely consistent with the current SFR ($L_{IR} \le L_{SF}$), indicating the IR emission is not due to current AGN activity. After removing the contribution of star formation to the IR luminosity, 5/6 post-starburst galaxies show evidence for fading AGN activity, with IR luminosities or limits less than expected if the AGN remained at the luminosity required to ionize the EELR.

\section{Discussion}

\subsection{Current AGN indicators}

Below, we consider several tracers of the current AGN activity in the sample of galaxies. We explore tracers of AGN activity using optical emission lines, IR colors, optical variability, and X-ray luminosity to compare the current AGN properties of the post-starburst galaxies in the MaNGA sample with those of the galaxies with EELRs.

\subsubsection{Optical Emission Line Ratios}

\begin{figure*}
\begin{center}
\includegraphics[width=0.49\textwidth]{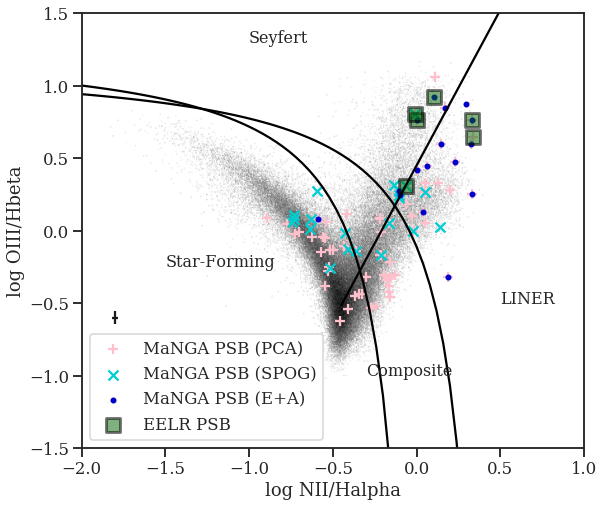}
\includegraphics[width=0.49\textwidth]{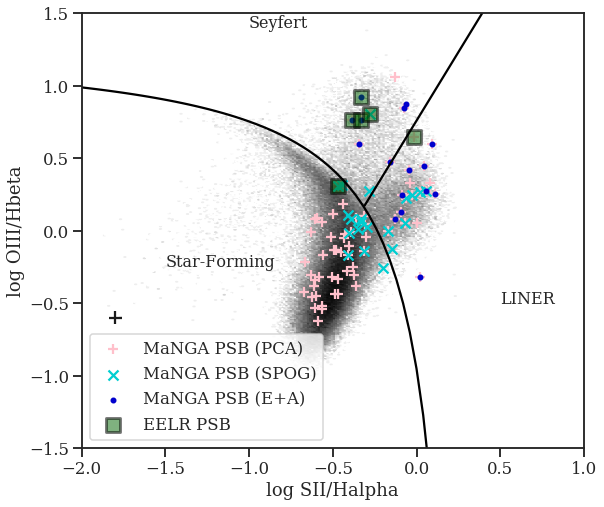}
\end{center}
\caption{Emission line ratio diagnostics for post-starburst galaxies and comparison galaxies from the SDSS main spectroscopic survey. Emission line fluxes from the {\tt galspec} catalogs \citep{Kauffmann2003, Brinchmann2004, Tremonti2004} are used for all galaxies shown. \textit{Left:} \nii/\halpha vs. \oiii/\hbeta diagram \citep{Baldwin1981}, with dividing lines from \citet{Kewley2001, Kauffmann2003}. \textit{Right:} \sii/\halpha vs. \oiii/\hbeta diagram \citep{veilleux87}, with dividing lines from \citet{Kewley2001, Kauffmann2003}. The characteristic uncertainty for the post-starburst sample is shown above the legend. Most of the post-starburst galaxies with EELRs have current emission consistent with being Seyferts or LINERs. 
}
\label{fig:bpt}
\end{figure*}

We use optical emission line ratios \citep{veilleux87, Baldwin1981, Kewley2001, Kauffmann2003} to trace the impact of star formation vs. AGN activity on the ionization of gas in this sample of galaxies. In Figure \ref{fig:bpt}, we consider the \nii/\halpha or \sii/\halpha vs. \oiii/\hbeta emission line ratios. The ionization source varies across this sample, with post-starburst galaxies spanning the range of star-forming, composite, LINER, and Seyfert. For the sample of post-starburst galaxies in MaNGA, the observed distribution depends on the post-starburst selection method, with E+A galaxies more likely to be LINERs and PCA galaxies more likely to be star-forming. The EELR galaxies are located primarily in the Seyfert/LINER portion of each diagram, such that Seyfert post-starburst galaxies have a much higher chance at hosting an EELR than the general post-starburst galaxy sample. However, this may be caused by contamination from the EELR in the central 3\arcsec\ SDSS fiber.

\begin{figure}
\begin{center}
\includegraphics[width=0.5\textwidth]{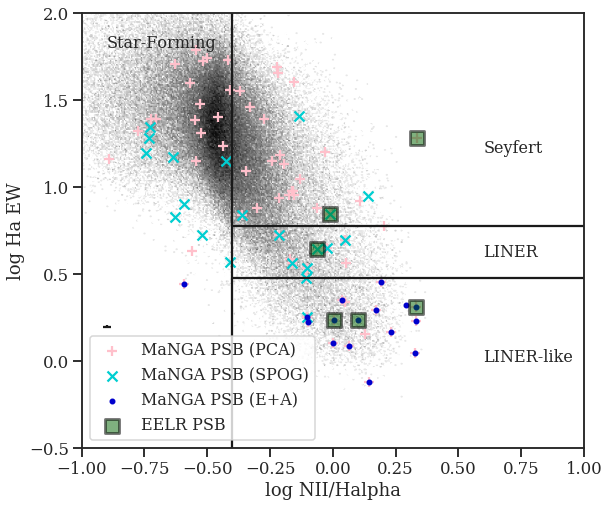}
\end{center}
\caption{Excitation diagram from \citet{CidFernandes2011} for the post-starburst galaxies and comparison galaxies from the SDSS main spectroscopic survey. Emission line fluxes and equivalent widths (EWs) from the {\tt galspec} catalogs \citep{Kauffmann2003a, Brinchmann2004, Tremonti2004} are used for all galaxies shown.  The characteristic uncertainty for the post-starburst sample is shown above the legend. The EELR galaxies span from two Seyferts, one LINER, and three LINER-like (or ``fake AGN") galaxies.
}
\label{fig:whan}
\end{figure}

Many of the post-starburst galaxies have weak emission lines, which may be caused by evolved stellar populations instead of true AGN activity. Because some of the post-starburst galaxy samples select against star formation, even faint sources of hard ionization can dominate over residual star formation. We consider the classification of the MaNGA post-starburst galaxies on a WHAN diagram \citep{CidFernandes2011} in Figure \ref{fig:whan}. The EELR galaxies span from two Seyferts, one LINER, and three LINER-like (or ``fake AGN") galaxies. Thus, the EELR sample likely contains a combination of Seyferts, low-luminosity AGN, and galaxies ionized by other sources of hard ionization. 

\subsubsection{Infrared AGN Selection from WISE}

\begin{figure}
\begin{center}
\includegraphics[width=0.49\textwidth]{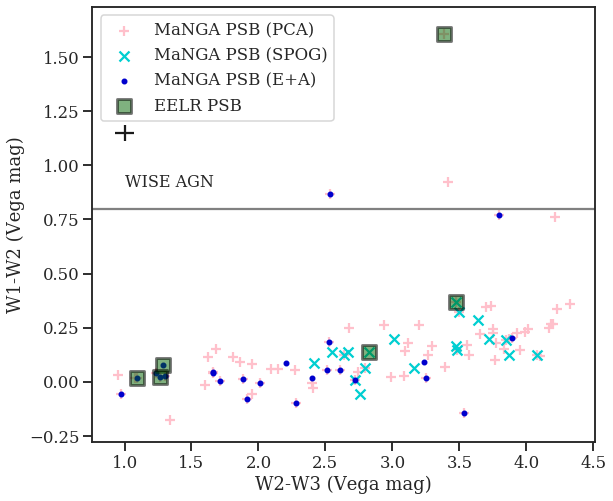}
\end{center}
\caption{WISE color-color plot, with \citet{Stern2012} AGN criterion overlaid. The characteristic uncertainty for the post-starburst sample is shown below the legend. Of our total sample of MaNGA post-starburst galaxies, 3/93 are WISE AGN; of the sample of post-starburst galaxies with EELRs, 1/6 (11965-3703)is a WISE AGN.
}
\label{fig:wise}
\end{figure}

Obscured AGN can be revealed by unique WISE colors \citep{Stern2012}. In Figure \ref{fig:wise}, we plot the WISE W1-W2 vs. W2-W3 colors. Galaxies with W1-W2$>0.8$ mag (Vega) show evidence for obscured AGN. Of our total sample of MaNGA post-starburst galaxies, 3/93 are WISE AGN; of the sample of post-starburst galaxies with EELRs, 1/6 is a WISE AGN. This galaxy (MaNGA plateifu 11965-3703) is a well-known AGN host that is also in the \citet{Keel2012} sample of voorwerpjes, and the only post-starburst galaxy EELR host detected in IRAS. 

\subsubsection{Optical Variability}

AGN emit radiation across the electromagnetic spectrum that varies stochastically in intensity on timescales ranging from hours to decades, with variability well-modeled by a damped random walk (DRW). This characteristic variability can be used to identify a potential AGN by searching for variability consistent with a DRW model. For each of the six post-starburst galaxies with EELRs, we obtain $r$, $g$, and $i$ band data from the Zwicky Transient Facility \citep[ZTF,][]{Bellm2019} to search for variable AGN signatures. We sigma clip the data in each band at the $5\sigma$ level twice, in order to remove outliers. We then fit each light curve using the DRW model of \citet{Butler2011}, which returns $\sigma_{\rm VAR}$, the significance of the source's variability, $\sigma_{\rm QSO}$, the significance of how DRW-like the source's variability is, and $\sigma_{\rm NOT \ QSO}$, the significance that the source's random variability is not AGN-like. To classify a source's variability as AGN-like, we require that the following be true in at least two of the three bands: $\sigma_{\rm VAR}$ $>$ 3, $\sigma_{\rm QSO}$ $>$ 3, and $\sigma_{\rm QSO}$ $>$ $\sigma_{\rm NOT \ QSO}$. We find that 5 of the 6 PSB EELRs have $\sigma_{\rm QSO}$ values less than 3, indicating that their variability is not AGN-like. The sixth galaxy, 9495-3702, returned high $\sigma_{\rm QSO}$ values of 12.94 and 29.82 in the r and g bands respectively, but it returned even higher $\sigma_{\rm NOT \ QSO}$ values of 72.70 and 33.97. The origin of this variability is the Tidal Disruption Event (TDE) AT2019azh (ASASSN-19dj, \citealt{Hinkle2021}). We explore the connection between TDEs and EELRs further in \S\ref{sec:tdes}. The lack of AGN-like variability in the six post-starburst galaxies with EELRs is consistent with the general lack of strong unobscured current AGN activity found by the tracers above, providing further evidence that these galaxies host fading AGN. Results for the full sample of $>5000$ post-starburst galaxies will be presented in Novack et al. (in prep).

\subsubsection{Current X-ray Luminosity}
\label{sec:xray}

One of the post-starburst galaxies with an EELR (9862-12703) has archival \textit{Chandra} observations (proposal ID: 10708322, PI Watson), which have been included in the \textit{Chandra} Source Catalog (CSC, \citealt{Evans2010}). We assume this X-ray emission is from an AGN, though much of the X-ray flux could be from a Quasi-Periodic Eruption \citep{Giustini2020}. In either scenario, the emission would be originating from the nucleus of this galaxy. Using {\tt sherpa}, we fit an absorbed power-law model to the X-ray spectrum for this galaxy to determine the current X-ray luminosity, $L_X = 2.2\pm 0.3\times10^{41}$ erg s$^{-1}$. This $L_X$ is consistent with the range of low X-ray luminosities in the sample of SPOGs considered by \citet{Lanz2022}. This galaxy has a current X-ray luminosity significantly lower than the current X-ray luminosity of the Hanny's voorwerp host galaxy, which has a current $L_x = 1-2.3\times10^{43}$ erg s$^{-1}$ \citep{Sartori2018b}.

We can compare the current X-ray luminosity for 9862-12703 with the past peak \lion as an alternative test to the use of the IR observations used above, to test whether this galaxy has a fading AGN. We use the calibrations of \citet{Marconi2004} to determine the bolometric luminosity implied by $L_X$, to be $\log L_{\rm bol}/(\rm erg \ s^{-1}) = 42.4$. Following \citet{Sartori2018b}, who use \citet{Elvis1994} models to estimate the bolometric correction from \lion, the lower limit on the past bolometric luminosity would be $\log L_{\rm bol}/(\rm erg \ s^{-1}) > 43.5$, significantly higher than what we infer from the current X-ray luminosity. 

\subsection{Observed Merger Signatures in Optical Photometry}
\label{sec:morphology}

\begin{figure*}
\begin{center}
\includegraphics[width=0.49\textwidth]{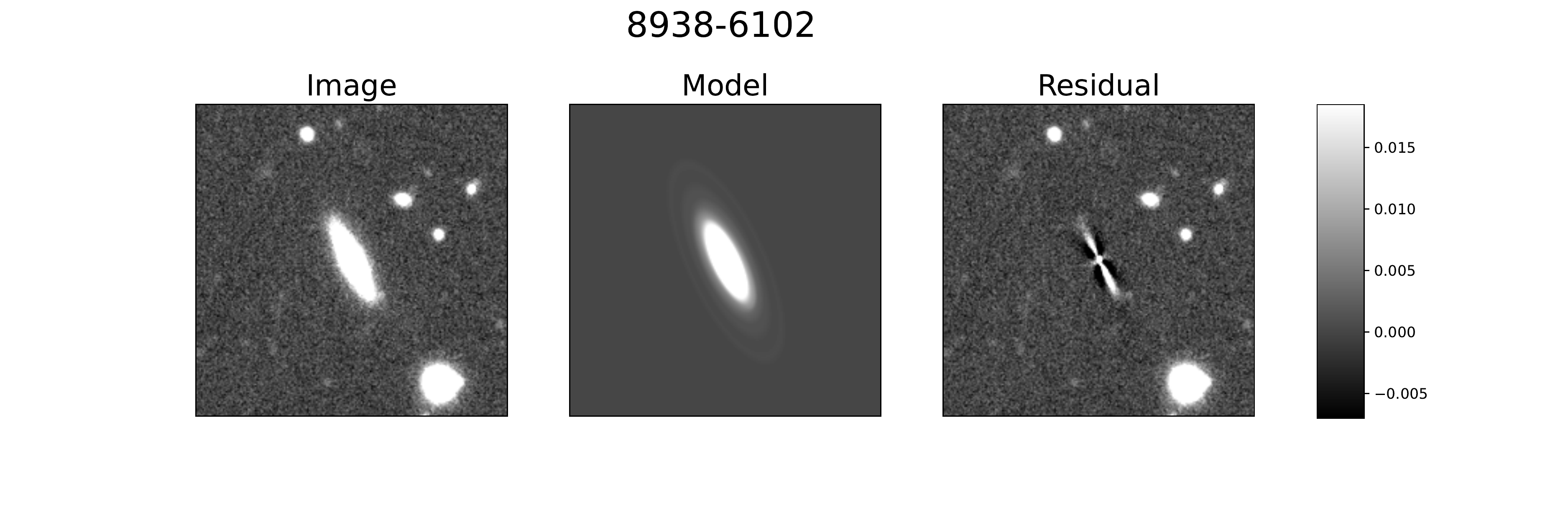}
\includegraphics[width=0.49\textwidth]{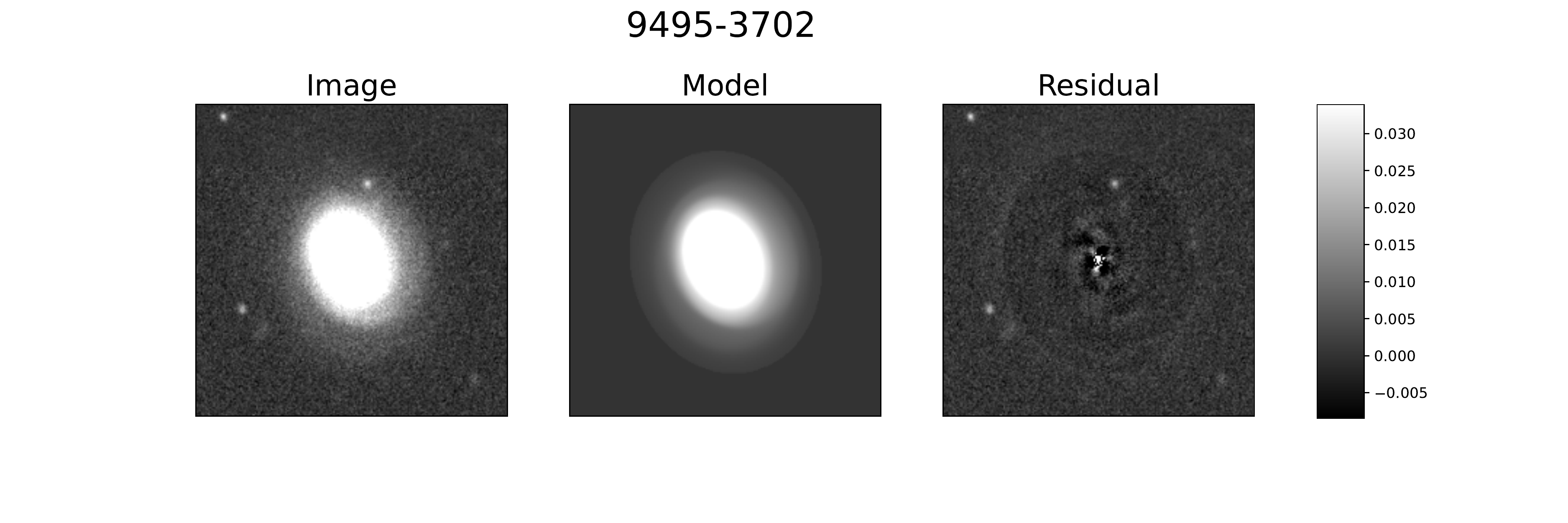}
\includegraphics[width=0.49\textwidth]{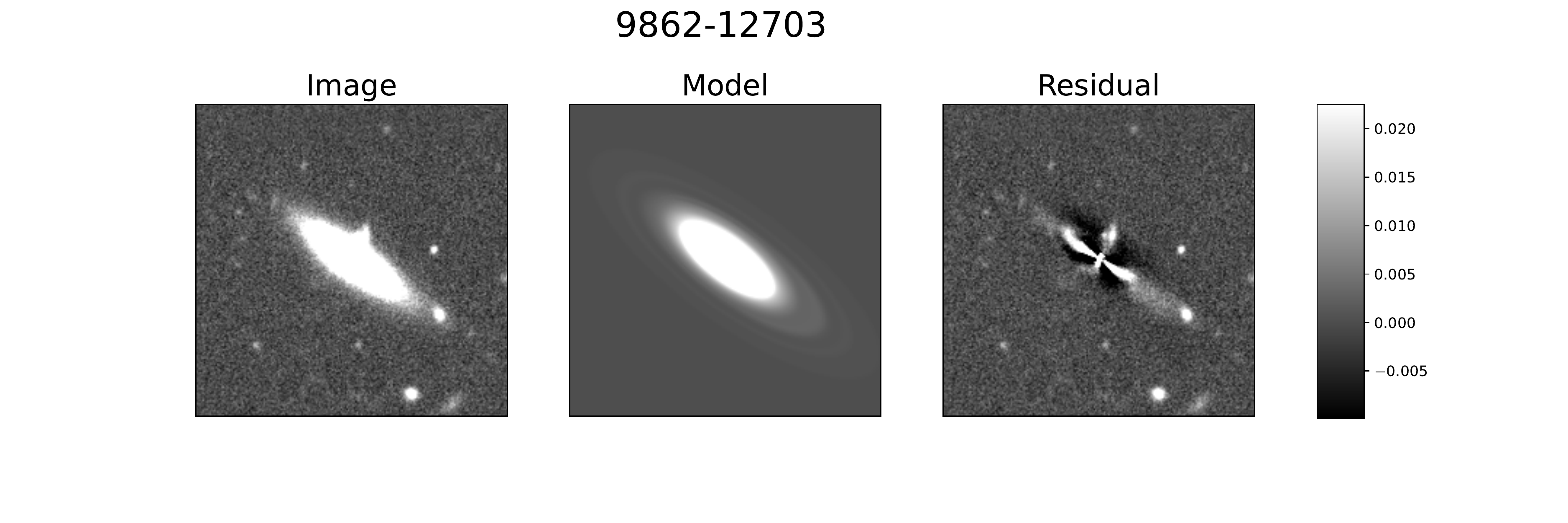}
\includegraphics[width=0.49\textwidth]{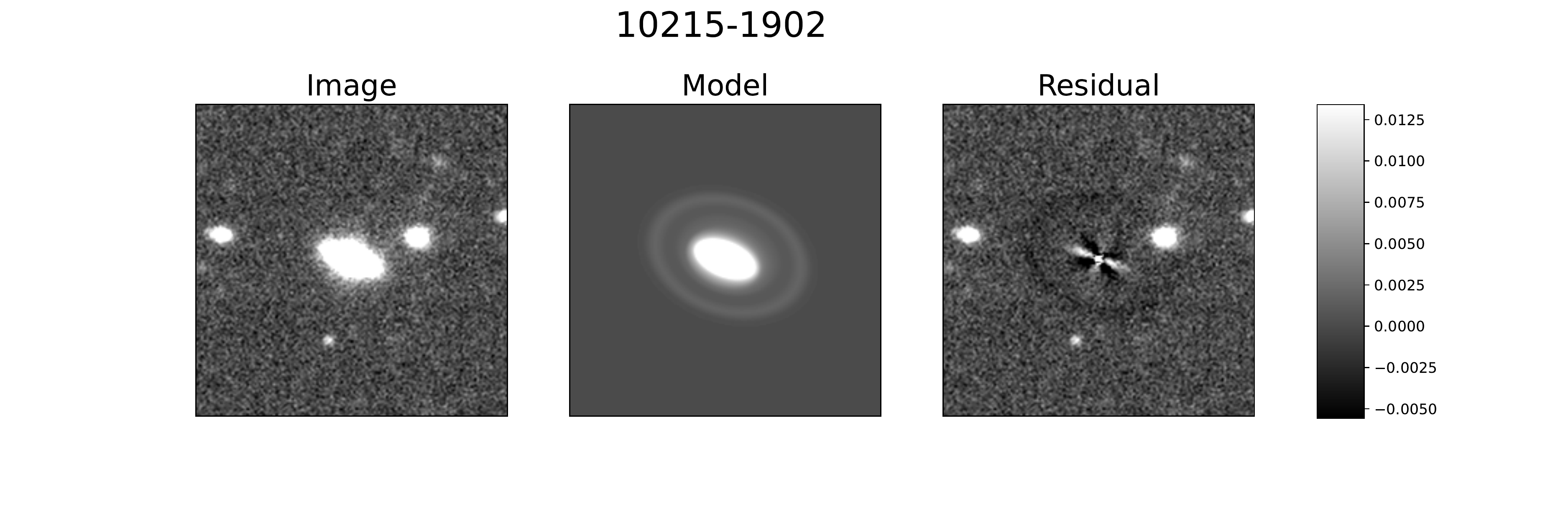}
\includegraphics[width=0.49\textwidth]{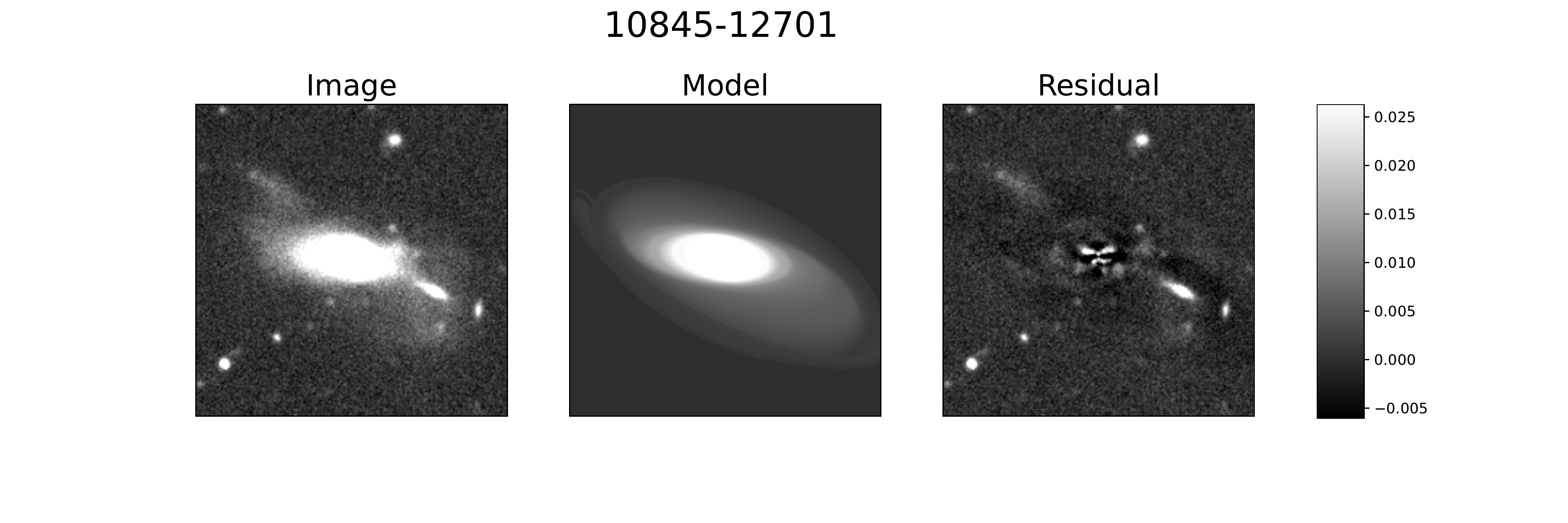}
\includegraphics[width=0.49\textwidth]{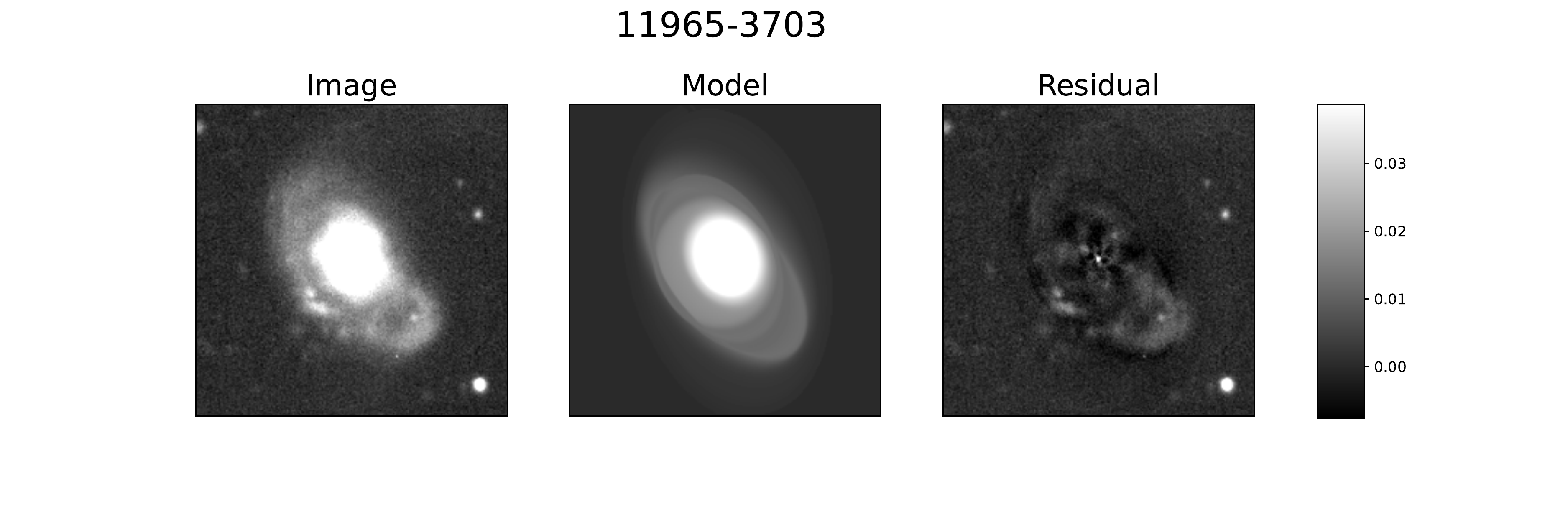}
\end{center}
\caption{Legacy survey g band images, elliptical disk models, and residual images for each of the six post-starburst galaxies with EELRs. Two/six (the bottom two cases shown here) show obvious signatures of a recent merger. A third galaxy shows a possible sign of interaction with a small satellite galaxy (9862-12703). The three remaining galaxies show no obvious signs of a recent merger, indicating that clear extended stellar features are not required in order for the galaxy to have sufficient extended gas to be illuminated in an EELR.
}
\label{fig:morph}
\end{figure*}

EELRs are more easily observed when a galaxy has sufficient extended gas to be ionized, so we may expect some or all of the post-starburst galaxies with EELRs to display signatures of recent galaxy-galaxy merging in their stellar light. To test this, we subtract off smooth models for each galaxy from the Legacy survey \citep{Dey2019} $g$-band imaging to examine the residuals for tidal tails and disturbed features. We use \texttt{photutils} ellipse fitting architecture to generate elliptical models for each galaxy.  First, we provide a guess for the semimajor axis, ellipticity, and position angle for each galaxy's $g$-band image, determined via visual inspection.  We then use \texttt{photutils} to fit elliptical isophotes out to a maximum semimajor axis, defined such that we avoid fitting other objects in the field.  Finally, we subtract the model from the initial image to generate a residual image.  The $g$-band images, elliptical models, and residuals are shown in Figure \ref{fig:morph}.

Two of the six galaxies (10845-12701, 11965-3703) show clear signs of disturbance in their stellar features. A third galaxy (9862-12703) has a nearby object which may be a satellite galaxy associated with the central galaxy, though we do not have a redshift to confirm. Because 4/6 of the galaxies do not show obvious merger features, we conclude that the presence of merger features in the stellar light is not necessary for the galaxy to have sufficient extended gas to display an EELR. These 4/6 galaxies may not have had a recent merger, or may have had a merger $\sim1$ Gyr ago, short enough such that we still see the post-starburst signature but long enough that merger features may have faded \citep{Pawlik2016}. Mergers are not the only method by which gas can be removed from galaxies; the fact that 3/4 of the undisturbed cases have disky morphologies viewed edge-on may be indicative of past biconical outflows, similar to those seen in the starburst galaxy M82, which are later ionized by luminous AGN activity. 

In order to measure a duty cycle of AGN activity in \S\ref{sec:dutycycle}, we must estimate the fraction of post-starburst galaxies capable of displaying an EELR. The EELR sample is not significantly more likely to contain edge-on disk galaxies (axis ratio b/a $<0.5$) than the total MaNGA post-starburst sample. Roughly half of post-starburst galaxies show signs of prominent tidal features \citep{Pracy2009, Yang2004, Yang2008, Pawlik2016}, with nearly 90\% having signs of disturbed morphologies on smaller scales \citep{Sazonova2021}. This is roughly consistent with the fraction of EELR post-starbursts with merger signatures. We thus use two bracketing estimates for the fraction of post-starburst galaxies capable of displaying an EELR: (1) we assume 100\% of the galaxies would display an EELR given sufficient recent AGN activity, since the fraction with mergers/edge-on disks is similar to the total populations, and (2) we assume 50\% of the galaxies would display an EELR given sufficient recent AGN activity, since at least this many galaxies have had a recent merger.

\subsection{Duty cycle of AGN activity during the post-starburst phase}
\label{sec:dutycycle}

With the fraction of observed EELRs in post-starburst galaxies and the timescale on which they are observable, we can estimate the duty cycle of strong AGN activity during this phase. The fraction of observed EELRs, $f$, is $f=6/93 = 0.065$. We estimate the typical observable EELR timescale from the typical time since peak \lion, $t_{\rm EELR}\sim1.5\times10^4$ years. If the remaining post-starburst galaxies have the same AGN duty cycle, yet have been dormant for $>1.5\times10^4$ years, this would imply a duty cycle $D = t_{\rm EELR} / f = 2.3\times 10^5 \ {\rm yr}$. However, some post-starburst galaxies may have recent AGN activity, but lack sufficient extended gas to result in observable EELRs. Using our bracketing estimate from \S\ref{sec:morphology}, we assume that at least half of the post-starburst galaxies would be capable of displaying EELRs given sufficient ionizing luminosity. This assumptions changes the fraction of observed EELRs to $f = 6/(93/2) = 0.13$, and our resulting duty cycle timescale to be $D = t_{\rm EELR} / f = 1.1\times 10^5 \  {\rm yr}$. 

Between the three tracers of current AGN activity discussed above, we assume we can see nearly all of the current strong AGN between the WISE AGN and the galaxies with Seyfert-like optical emission line ratios. Two of the 93 post-starburst galaxies are consistent with being Seyferts on all three of the \nii, \sii, and WHAN diagrams above\footnote{We use this strict cut in order to avoid contamination from hard ionization from shocks or evolved stars, both of which are likely to be common in post-starburst galaxies.}. Three additional of the 93 galaxies are WISE AGN (including one of the EELR hosts), for a total of 5/93 of the post-starburst galaxies showing evidence for current strong AGN. Assuming 6--12 of the 93 had recent AGN activity in the last $t_{\rm EELR}\sim1.5\times10^4$ years following the arguments above, we can combine the prevalence of past and current AGN activity to estimate the time the AGN is typically ``on" over each duty cycle to be roughly 5/93 (5.3\%) or $t_{\rm ON} = 0.6-1.3\times10^4$ yr. 

We compare the fraction of current AGN observed in the MaNGA post-starburst sample to the full MaNGA DR17 sample. Selecting current AGN using the same WISE and optical emission line cuts as described above, we find the AGN fraction to be 3.8\%, less than the 5.3\% found for the post-starburst galaxies, albeit with low significance given the small number of post-starburst AGN. The fraction of time AGN spend ``on" is expected to be low for the low redshift sample we consider here, $\sim0.01-1$\% \citep{Shankar2009}. Using optical line ratios to identify current AGN, \citet{Schawinski2010} find AGN fractions ranging from 0.1-10\% depending on galaxy type. The galaxies with the largest AGN fraction were green valley early type galaxies with stellar masses $\sim10^{9.5}-10^{10}$ \Msun, which are similar to the post-starburst galaxy EELR hosts. \citet{Pawlik2018} previously estimated the fraction of current AGN in post-starburst galaxies to be 50\%, but this estimate is based on a less-strict cut on optical emission line ratios, and may be contaminated by galaxies with ionization dominated by shocks or evolved stars. Our estimate for this post-starburst sample is consistent with the $\sim5$\% AGN fraction found by \citet{Greene2020} for a sample of $z\sim0.7$ post-starburst galaxies.

\subsection{Context within the post-starburst galaxy population}

\begin{figure*}
\begin{center}
\includegraphics[width=0.49\textwidth]{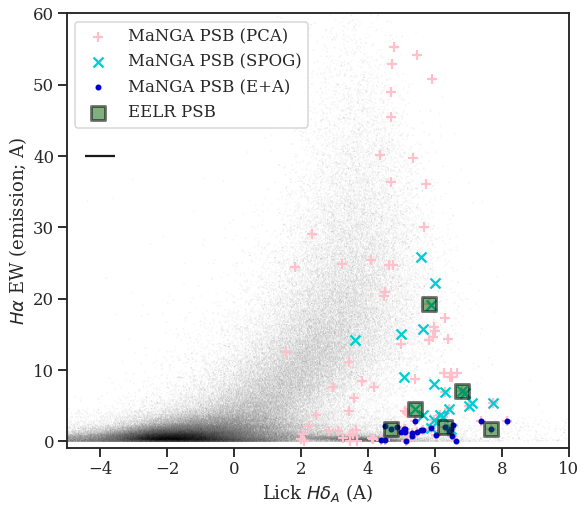}
\includegraphics[width=0.49\textwidth]{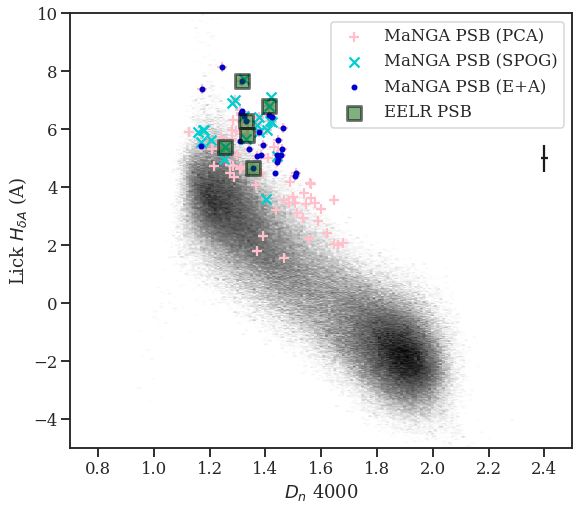}
\end{center}
\caption{Tracers of star formation history in post-starburst galaxies and those with EELRs. Comparison galaxies from {\tt galspec} are shown in greyscale. \textit{Left:} Lick \hda vs. \halpha EW, the space used to select the E+A subsample of post-starburst galaxies. The \halpha EW traces the current SFR, which \hda traces star formation over the last $\sim$Gyr. The characteristic uncertainty for the post-starburst sample is shown below the legend for each panel. The E+A galaxies all lie in the corner of this diagram by selection, though the other post-starburst samples also have strong H$\delta$ absorption. The post-starburst galaxies with EELRs tend to have stronger \hda absorption. 
\textit{Right:} $D_n(4000)$ vs. Lick \hda, related to the space used to select the PCA subsample. $D_n(4000)$ traces the average stellar age of the galaxy, with star-forming galaxies having lower $D_n(4000)$ than quiescent galaxies. The Lick \hda absorption, when measured relative to this average trend in $D_n(4000)$, traces how bursty the recent star-formation has been. 
}
\label{fig:hahd}
\end{figure*}

We consider the stellar populations of the sample of MaNGA post-starburst galaxies and the subsample with EELRs using spectroscopic tracers in Figure \ref{fig:hahd}. Figure \ref{fig:hahd}a shows the current star formation traced by \halpha emission against star formation over the last $\sim$ Gyr traced by the Lick \hda index. The E+A galaxies all lie in the corner of this diagram by selection, though the other post-starburst samples also have strong H$\delta$ absorption. Figure \ref{fig:hahd}b shows the same Lick \hda index, now against the $D_n(4000)$ break index. This space is similar to the one used to select the PCA sample, and all of the post-starburst galaxies lie at the bursty end of this sequence \citep{Kauffmann2003a}. The EELR galaxies lie at higher Lick \hda. A Mann-Whitney U test shows that the parent post-starburst population and EELR population are drawn from different distributions of \hda at the $2\sigma$ significance level.

\begin{figure}
\begin{center}
\includegraphics[width=0.49\textwidth]{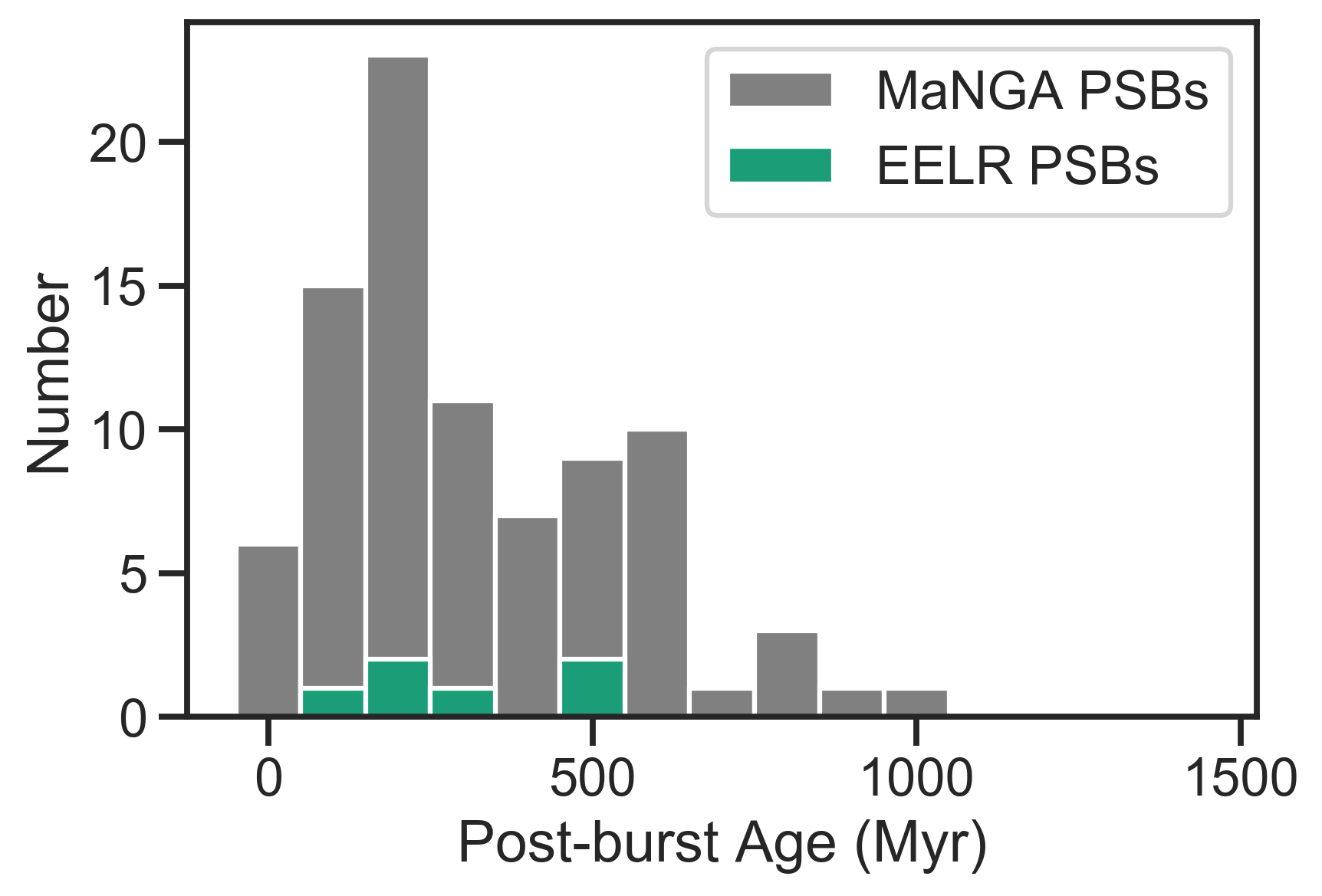}
\end{center}
\caption{Distribution of post-burst ages of the full MaNGA post-starburst sample (grey) and post-starburst galaxies with EELRs (green). We observe the post-burst ages of the post-starburst galaxies with EELRs to span a wide range, from 140--520 Myr. We see no evidence that these galaxies are drawn from a different population in age than the parent MaNGA post-starburst sample. This indicates that AGN activity may act throughout the post-starburst phase. 
}
\label{fig:age}
\end{figure}

We test whether the higher \hda is caused by a difference in post-burst age in Figure \ref{fig:age}. We obtain post-burst ages from the stellar population fits done by \citet{French2018}. For galaxies not included in that work, we obtain the ages using the same methodology, using the SDSS spectral indices along with \textit{GALEX} and SDSS photometry to fit a set of parametric star formation histories. We observe the post-burst ages of the post-starburst galaxies with EELRs to span a wide range, from 140--520 Myr. We see no evidence that these galaxies are drawn from a different population in age than the parent MaNGA post-starburst sample. This indicates that AGN activity may act throughout the post-starburst phase. 

Higher \hda in post-starburst galaxies could be caused by either stronger starbursts (higher fractions of stellar mass produced during the last Gyr), the post-starburst ages being younger, or the duration of the starburst being shorter \citep[see e.g.,][]{French2018}. We also observe no evidence that the EELR galaxies have differing burst mass fractions or burst durations. The trend of the EELR post-starbursts to have higher \hda may be spurious or due to a more complex correlation with these starburst properties.

\begin{figure}
\begin{center}
\includegraphics[width=0.49\textwidth]{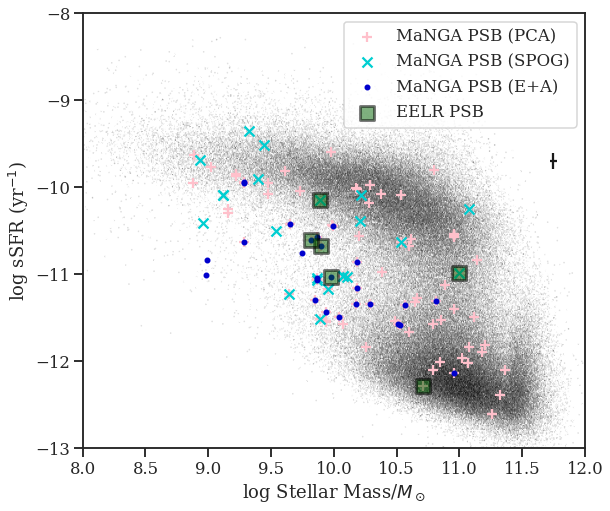}
\end{center}
\caption{sSFR vs. stellar mass for the post-starburst galaxies and comparison galaxies from the SDSS main spectroscopic survey. The characteristic uncertainty for the post-starburst sample is shown below the legend, though the true uncertainties will likely be dominated by systematic effects. The post-starburst galaxies span across this space, from star-forming to green valley to quiescent, with most consistent with the green valley. The post-starburst galaxies with EELRs are similarly distributed, suggesting that any AGN feedback occurring in these galaxies has not acted to quickly shut down star formation.
}
\label{fig:sfms}
\end{figure}

The specific SFR (sSFR) vs. stellar mass for the post-starburst galaxies and comparison galaxies from the SDSS main spectroscopic survey are shown in Figure \ref{fig:sfms}. Star formation rates and stellar masses for all galaxies are from the MPA-JHU galspec catalogs \citep{Kauffmann2003a, Tremonti2004, Brinchmann2004, Salim2007} and utilize H$\alpha$ fluxes, $D_n(4000)$ indices, and galaxy colors in order to obtain SFRs. These SFRs are corrected for dust attenuation using the Balmer decrement, but may miss the presence of heavily embedded star formation (see further discussions in \citealt{Smercina2022, Baron2022b, French2023}). The post-starburst galaxies span across this space, from star-forming to green valley to quiescent, with most consistent with the green valley. The post-starburst galaxies with EELRs are similarly distributed, suggesting that any AGN feedback occurring in these galaxies has not acted to quickly shut down star formation. This is consistent with the picture from \S\ref{sec:dutycycle}, suggesting that AGN may act over multiple cycles instead of in a single rapid blowout.

\subsection{Ionized Gas Kinematics}

The MaNGA data analysis pipeline team \citep{Westfall2019} has constructed maps of the stellar and ionized gas kinematics of each galaxy. For the six post-starburst galaxies with EELRs, 5/6 have clear misalignments between the ionized gas and stellar velocity. Most of the galaxies show some stellar rotation (further analysis of the stellar kinematics of this sample will be presented in Pardasani et al. in prep). The ionized gas kinematics show a range of properties with some (8938-6102, 9495-3702) showing evidence of rotation and others (10845-12701, 11965-3703) displaying asymmetric biconical patterns. These may be evidence of outflows, as indeed 10845-12701 is known to host an AGN outflow \citep{McElroy2015}. In one case (11965-3703), the ionized gas velocity red/blue components are aligned with the orientation of the EELR; in another (10845-12701), the ionized gas velocities are oriented $\sim45$ degrees from the EELR. Future study of the kinematics of the ionized gas as well as other gas phases will yield useful constraints on the impact of past AGN activity on each galaxy.

\subsection{Connection to Other Forms of Nuclear Activity}
\label{sec:tdes}

Several of the post-starburst galaxies with EELRs have evidence for recent nuclear activity likely caused by sources other than a persistent AGN gas disk. 9495-3702 is the host galaxy of the Tidal Disruption Event (TDE) AT2019azh/ASASSN-19dj \citep{Hinkle2021}. The observed optical, UV, and X-ray flare in 2019 was driven by the accretion of debris from an individual star after the star was tidally disrupted by the SMBH. This event was too recent to have ionized the EELR. A second post-starburst galaxy with an EELR, 9862-12703, is the host galaxy to one of the few observed Quasi Periodic Eruption (QPE) sources \citep{Giustini2020}. The physical origin of QPEs is unknown, but may be related to extreme mass ratio inspirals or tidal disruptions of white dwarfs \citep[e.g.,][]{King2020, Metzger2022, Wang2022}.

Six TDE host galaxies meet a strict E+A cut to be considered post-starburst: PTF09axc, PTF09djl, ASASSN-14li, iPTF16fnl, AT2018hyz (ASASSN-18zj), and AT2019azh (ASASSN-19dj) \citep{Arcavi2014, French2016, Graur2018, French2020, Hinkle2021}. Of these, 2/6 have been observed using integral field spectroscopy: AT2019azh (considered here) and ASASSN-14li \citep{Prieto2016}, with both host galaxies displaying EELRs. Given the low rate of EELRs in the post-starburst sample, it is unusual to see two such cases in the post-starburst TDE host subsample. QPE host galaxies show a similar overrepresentation in post-starburst galaxies as TDEs \citep{Wevers2022}, and similarly, the only known post-starburst QPE host is a post-starburst EELR galaxy. One possibility to explain this connection is if the TDE rate is increased by previous interactions with an AGN disk \citep{Kennedy2016}. This effect may act in concert with selection biases against claiming TDE discoveries in known AGN, such that the post-starburst EELR galaxies are cases where the stellar orbits could have been affected by interactions with an AGN disk yet the AGN has faded such that AGN variability is no longer a concern in detecting TDEs. If the TDE host galaxies are more likely to be not just post-starburst galaxies but post-starburst galaxies with EELRs, this implies that the TDE rate enhancement is $\sim2-15\times$ higher than previous estimates for TDE rate enhancement in post-starburst galaxies.

\subsection{Implications for SMBH growth and feedback}

Using the peak AGN luminosities and duty cycles determined above, we can estimate the SMBH mass growth over the post-starburst phase. This newly added mass can be estimated as
\begin{equation}
    M_{\rm new} = \frac{\Delta t L_{AGN}}{\epsilon c^2}
\end{equation}
where $\Delta t$ is the total time the AGN spends ``on", $L_{AGN}$ is the luminosity of the AGN during the these periods, and $\epsilon$ is the efficiency of conversion of accreted mass to luminosity. In order to determine $\Delta t$, we take the typical post-burst age of the EELR sample ($t \sim 300$ Myr) and multiply by the fraction of time we estimate the AGN to be on (5.3\% from above). We use the ionizing luminosity \lion and the estimate from \S\ref{sec:xray} that the typical ratio of \lion to the bolometric luminosity $L_{AGN}$ is 0.2 to estimate a typical $L_{AGN} \sim 10^{43.6}-10^{45.2}$ erg s$^{-1}$. We assume an accretion efficiency of $\epsilon = 0.1$. Our resulting newly added mass estimate is $M_{\rm new} = 1.1\times10^5-5.0\times10^6$ \Msun. For an estimated SMBH mass in these galaxies of $M_{\rm BH} = 10^7$ \Msun, this a SMBH growth of 1.1--50\% during the post-starburst phase. This growth is broadly consistent with the 5\% growth measurement from \citet{Wild2010} on a PCA-selected sample of post-starburst galaxies.

Is this energy enough to affect the ability of these galaxies to form stars? The total energy released by the SMBH during the post-starburst phase using the estimate above is $E_{\rm tot} = 2.0 \times 10^{58}$ erg. We can estimate the binding energy of the molecular gas in these galaxies using the typical molecular gas mass of $M_{\rm gas} = 10^9$ \Msun and gas FWHM of $200$ km s$^{-1}$ from \citet{French2015}. The binding energy of the gas can be estimated using $E_{\rm gas} \sim M_{\rm gas} \sigma^2 = 1.4\times10^{56}$ erg. Thus, even considering the lower bound of total mass accreted onto the SMBH during this, the energy is still a factor of $\sim140\times$ more than what would be needed to unbind the molecular gas assuming perfect coupling of the energy. The total AGN energy released during the post-starburst phase is somewhat lower than the total energy observed in simulated galaxies by \citet{Lotz2021}, where a total energy of $E_{\rm tot} > 6 \times 10^{58}$ erg would be released over 600 Myr. 

While our results here do not provide constraints on the mechanisms that ended the recent starbursts $\sim 300$ Myr ago, the characterization of these EELRs is consistent with the picture \citep[proposed by e.g.,][]{Davies2007, Schawinski2009, Wild2010, Hopkins2012b, Cales2015} that some AGN feedback is delayed after the starburst. This delayed feedback is capable of depleting the molecular gas supply that persists into the post-starburst phase \citep{French2015, Rowlands2015, Alatalo2016b}, depending on how the energy is coupled to the gas. It remains unclear why the low-level AGN activity we infer here might have an effect on driving these galaxies to quiescence, yet luminous AGN are often observed to reside in steadily star-forming galaxies \citep[e.g.,][]{Florez2020}. The disturbance to the galaxy caused by the recent merger and/or starburst may act to make post-starburst galaxies more susceptible to AGN feedback \citep{Pontzen2017, Davies2022}. Understanding the nature of how AGN energy couples to the ISM, and whether the efficiency of this coupling varies across the galaxy population, will be crucial to understanding the role of AGN feedback in galaxy evolution.

\section{Conclusions}

We use integral field spectroscopy from MaNGA for a sample of 93 post-starburst galaxies to search for evidence of past AGN activity. Six post-starburst galaxies show EELRs, analogous to luminous ``voorwerps" seen in optical imaging. Our main conclusions are as follows:

\begin{enumerate}
    \item Of the post-starburst galaxies with EELRs, 5/6 are likely fading AGN, with current AGN luminosities lower than the peak minimum luminosity required to have ionized the EELR. The past ionizing luminosities of these 6 galaxies are at least Seyfert-like, with \lion$\gtrsim10^{42.5}-10^{44}$ erg s$^{-1}$. Our constraints on the upper limit of past \lion range from $10^{44.2}-10^{45.3}$. 

    \item The past AGN luminosities are in contrast with the often weak indicators of current AGN activity. Only 3/6 of the post-starburst galaxies with EELRs are Seyfert- or LINER-like on a WHAN \citep{CidFernandes2011} diagram; most are consistent with the Seyfert/LINER portion of optical emission line ratio diagrams, yet with weak optical line fluxes suggesting the current ionization source may be instead shocks or evolved stars. Similarly, only 1/6 of the post-starburst galaxies with EELRs is a WISE AGN and none of the galaxies show optical variability. When compared to the full sample of post-starburst galaxies in MaNGA, those with Seyfert-like ionization are more likely to host a EELR than the general population of post-starbursts, though this may be caused by contamination from the EELR in the central 3\arcsec\ SDSS fiber.
    
    \item Given the rate at which we observe EELRs, the typical EELR visibility timescale, and an estimate of how often EELRs would be visible, we estimate the duty cycle of AGN activity during the post-starburst phase. The timescale for the galaxy to cycle between peaks in AGN luminosity is $t_{\rm EELR}\sim1.1-2.3\times10^5$ yr. Given the rate at which we observe current AGN activity during this phase, we estimate that the AGN spends only 5.3\% of this time (or $t_{\rm ON} = 0.6-1.3\times10^4$ yr) in its luminous phase, with the rest of the time spent ``off" or in a low-luminosity phase. The length of this duty cycle may explain why so few luminous AGN have been observed during the post-starburst phase, despite evidence for AGN feedback at work.

    \item The post-starburst galaxies with EELRs occupy a similar range in post-starburst age and sSFR as the rest of the post-starburst galaxy population, suggesting that AGN activity persists throughout the post-starburst phase and that any AGN feedback occurring in these galaxies has not acted to quickly shut down star formation. 

    \item The SMBH growth we infer during the post-starburst phase ranges from $\sim1.1-50$\% for a typical SMBH mass of $M_{\rm BH} = 10^7$ \Msun. Understanding how this energy couples to the gas in these galaxies will be crucial to understanding the role AGN feedback may play in quenching galaxies and evolving them to quiescence. 
\end{enumerate}

These systems will be valuable targets for future modeling of the full emission line spectra to determine the past AGN luminosity beyond the simple recombination balance method presented here, as well as for determining the impact of the AGN on the kinematics of the gas in ionized, neutral, and molecular phases. Future deep imaging surveys (especially the Vera C. Rubin Observatory Legacy Survey of Space and Time) will provide more opportunities to discover EELRs in post-starburst galaxies where the contribution of the strong line emission is detectable in the deep broadband imaging. 

\vspace{0.5cm}

We thank the referee for useful feedback, which has improved this manuscript. This publication makes use of data products from the Wide-field Infrared Survey Explorer, which is a joint project of the University of California, Los Angeles, and the Jet Propulsion Laboratory/California Institute of Technology, funded by the National Aeronautics and Space Administration.

This research has made use of data obtained from the Chandra Source Catalog, provided by the Chandra X-ray Center (CXC) as part of the Chandra Data Archive.

Funding for the Sloan Digital Sky 
Survey IV has been provided by the 
Alfred P. Sloan Foundation, the U.S. 
Department of Energy Office of 
Science, and the Participating 
Institutions. 

SDSS-IV acknowledges support and 
resources from the Center for High 
Performance Computing  at the 
University of Utah. The SDSS 
website is www.sdss.org.

SDSS-IV is managed by the 
Astrophysical Research Consortium 
for the Participating Institutions 
of the SDSS Collaboration including 
the Brazilian Participation Group, 
the Carnegie Institution for Science, 
Carnegie Mellon University, Center for 
Astrophysics | Harvard \& 
Smithsonian, the Chilean Participation 
Group, the French Participation Group, 
Instituto de Astrof\'isica de 
Canarias, The Johns Hopkins 
University, Kavli Institute for the 
Physics and Mathematics of the 
Universe (IPMU) / University of 
Tokyo, the Korean Participation Group, 
Lawrence Berkeley National Laboratory, 
Leibniz Institut f\"ur Astrophysik 
Potsdam (AIP),  Max-Planck-Institut 
f\"ur Astronomie (MPIA Heidelberg), 
Max-Planck-Institut f\"ur 
Astrophysik (MPA Garching), 
Max-Planck-Institut f\"ur 
Extraterrestrische Physik (MPE), 
National Astronomical Observatories of 
China, New Mexico State University, 
New York University, University of 
Notre Dame, Observat\'ario 
Nacional / MCTI, The Ohio State 
University, Pennsylvania State 
University, Shanghai 
Astronomical Observatory, United 
Kingdom Participation Group, 
Universidad Nacional Aut\'onoma 
de M\'exico, University of Arizona, 
University of Colorado Boulder, 
University of Oxford, University of 
Portsmouth, University of Utah, 
University of Virginia, University 
of Washington, University of 
Wisconsin, Vanderbilt University, 
and Yale University.

This research has made use of the NASA/IPAC Infrared Science Archive, which is funded by the National Aeronautics and Space Administration and operated by the California Institute of Technology.

\software{Astropy \citep{astropy2013, astropy2018}, Matplotlib \citep{matplotlib}, NumPy \citep{numpy}, SDSS-Marvin \citep{Cherinka19}}

\facility{IRSA, WISE, IRAS, Chandra}

Note: authors after first are listed in alphabetical order.


\bibliographystyle{aasjournal}
\bibliography{references}

\begin{thebibliography}{}
\expandafter\ifx\csname natexlab\endcsname\relax\def\natexlab#1{#1}\fi
\providecommand{\url}[1]{\href{#1}{#1}}
\providecommand{\dodoi}[1]{doi:~\href{http://doi.org/#1}{\nolinkurl{#1}}}
\providecommand{\doeprint}[1]{\href{http://ascl.net/#1}{\nolinkurl{http://ascl.net/#1}}}
\providecommand{\doarXiv}[1]{\href{https://arxiv.org/abs/#1}{\nolinkurl{https://arxiv.org/abs/#1}}}

\bibitem[{Alatalo {et~al.}(2016{\natexlab{a}})Alatalo, Cales, Rich, Appleton,
  Kewley, Lacy, Lanz, Medling, \& Nyland}]{Alatalo2016a}
Alatalo, K., Cales, S.~L., Rich, J.~A., {et~al.} 2016{\natexlab{a}}, \apjs,
  224, 38, \dodoi{10.3847/0067-0049/224/2/38}

\bibitem[{Alatalo {et~al.}(2016{\natexlab{b}})Alatalo, Lisenfeld, Lanz,
  Appleton, Cales, Kewley, Lacy, Medling, Nyland, Rich, Urry, \&
  Urry}]{Alatalo2016b}
Alatalo, K., Lisenfeld, U., Lanz, L., {et~al.} 2016{\natexlab{b}}, \apj, 827,
  106, \dodoi{10.3847/0004-637X/827/2/106}

\bibitem[{Arcavi {et~al.}(2014)Arcavi, Gal-Yam, Sullivan, Pan, Cenko, Horesh,
  Ofek, {De Cia}, Yan, Yang, Howell, Tal, Kulkarni, Tendulkar, Tang, Xu,
  Sternberg, Cohen, Bloom, Nugent, Kasliwal, Perley, Quimby, Miller, Theissen,
  \& Laher}]{Arcavi2014}
Arcavi, I., Gal-Yam, A., Sullivan, M., {et~al.} 2014, \apj, 793, 38,
  \dodoi{10.1088/0004-637X/793/1/38}

\bibitem[{{Ardila} {et~al.}(2018){Ardila}, {Alatalo}, {Lanz}, {Appleton},
  {Beaton}, {Bitsakis}, {Cales}, {Falc{\'o}n-Barroso}, {Kewley}, {Medling},
  {Mulchaey}, {Nyland}, {Rich}, \& {Urry}}]{Ardila2018}
{Ardila}, F., {Alatalo}, K., {Lanz}, L., {et~al.} 2018, \apj, 863, 28,
  \dodoi{10.3847/1538-4357/aad0a3}

\bibitem[{{Astropy Collaboration} {et~al.}(2013){Astropy Collaboration},
  {Robitaille}, {Tollerud}, {Greenfield}, {Droettboom}, {Bray}, {Aldcroft},
  {Davis}, {Ginsburg}, {Price-Whelan}, {Kerzendorf}, {Conley}, {Crighton},
  {Barbary}, {Muna}, {Ferguson}, {Grollier}, {Parikh}, {Nair}, {Unther},
  {Deil}, {Woillez}, {Conseil}, {Kramer}, {Turner}, {Singer}, {Fox}, {Weaver},
  {Zabalza}, {Edwards}, {Azalee Bostroem}, {Burke}, {Casey}, {Crawford},
  {Dencheva}, {Ely}, {Jenness}, {Labrie}, {Lim}, {Pierfederici}, {Pontzen},
  {Ptak}, {Refsdal}, {Servillat}, \& {Streicher}}]{astropy2013}
{Astropy Collaboration}, {Robitaille}, T.~P., {Tollerud}, E.~J., {et~al.} 2013,
  \aap, 558, A33, \dodoi{10.1051/0004-6361/201322068}

\bibitem[{{Astropy Collaboration} {et~al.}(2018){Astropy Collaboration},
  {Price-Whelan}, {Sip{\H{o}}cz}, {G{\"u}nther}, {Lim}, {Crawford}, {Conseil},
  {Shupe}, {Craig}, {Dencheva}, {Ginsburg}, {VanderPlas}, {Bradley},
  {P{\'e}rez-Su{\'a}rez}, {de Val-Borro}, {Aldcroft}, {Cruz}, {Robitaille},
  {Tollerud}, {Ardelean}, {Babej}, {Bach}, {Bachetti}, {Bakanov}, {Bamford},
  {Barentsen}, {Barmby}, {Baumbach}, {Berry}, {Biscani}, {Boquien}, {Bostroem},
  {Bouma}, {Brammer}, {Bray}, {Breytenbach}, {Buddelmeijer}, {Burke},
  {Calderone}, {Cano Rodr{\'\i}guez}, {Cara}, {Cardoso}, {Cheedella}, {Copin},
  {Corrales}, {Crichton}, {D'Avella}, {Deil}, {Depagne}, {Dietrich}, {Donath},
  {Droettboom}, {Earl}, {Erben}, {Fabbro}, {Ferreira}, {Finethy}, {Fox},
  {Garrison}, {Gibbons}, {Goldstein}, {Gommers}, {Greco}, {Greenfield},
  {Groener}, {Grollier}, {Hagen}, {Hirst}, {Homeier}, {Horton}, {Hosseinzadeh},
  {Hu}, {Hunkeler}, {Ivezi{\'c}}, {Jain}, {Jenness}, {Kanarek}, {Kendrew},
  {Kern}, {Kerzendorf}, {Khvalko}, {King}, {Kirkby}, {Kulkarni}, {Kumar},
  {Lee}, {Lenz}, {Littlefair}, {Ma}, {Macleod}, {Mastropietro}, {McCully},
  {Montagnac}, {Morris}, {Mueller}, {Mumford}, {Muna}, {Murphy}, {Nelson},
  {Nguyen}, {Ninan}, {N{\"o}the}, {Ogaz}, {Oh}, {Parejko}, {Parley}, {Pascual},
  {Patil}, {Patil}, {Plunkett}, {Prochaska}, {Rastogi}, {Reddy Janga},
  {Sabater}, {Sakurikar}, {Seifert}, {Sherbert}, {Sherwood-Taylor}, {Shih},
  {Sick}, {Silbiger}, {Singanamalla}, {Singer}, {Sladen}, {Sooley},
  {Sornarajah}, {Streicher}, {Teuben}, {Thomas}, {Tremblay}, {Turner},
  {Terr{\'o}n}, {van Kerkwijk}, {de la Vega}, {Watkins}, {Weaver}, {Whitmore},
  {Woillez}, {Zabalza}, \& {Astropy Contributors}}]{astropy2018}
{Astropy Collaboration}, {Price-Whelan}, A.~M., {Sip{\H{o}}cz}, B.~M., {et~al.}
  2018, \aj, 156, 123, \dodoi{10.3847/1538-3881/aabc4f}

\bibitem[{{Baldry} {et~al.}(2004){Baldry}, {Glazebrook}, {Brinkmann},
  {Ivezi{\'c}}, {Lupton}, {Nichol}, \& {Szalay}}]{Baldry2004}
{Baldry}, I.~K., {Glazebrook}, K., {Brinkmann}, J., {et~al.} 2004, \apj, 600,
  681, \dodoi{10.1086/380092}

\bibitem[{{Baldwin} {et~al.}(1981){Baldwin}, {Phillips}, \&
  {Terlevich}}]{Baldwin1981}
{Baldwin}, J.~A., {Phillips}, M.~M., \& {Terlevich}, R. 1981, \pasp, 93, 5,
  \dodoi{10.1086/130766}

\bibitem[{{Baron} {et~al.}(2022{\natexlab{a}}){Baron}, {Netzer}, {French},
  {Lutz}, {Davies}, \& {Prochaska}}]{Baron2022b}
{Baron}, D., {Netzer}, H., {French}, K.~D., {et~al.} 2022{\natexlab{a}}, arXiv
  e-prints, arXiv:2204.11881.
\newblock \doarXiv{2204.11881}

\bibitem[{{Baron} {et~al.}(2022{\natexlab{b}}){Baron}, {Netzer}, {Lutz},
  {Prochaska}, \& {Davies}}]{Baron2022a}
{Baron}, D., {Netzer}, H., {Lutz}, D., {Prochaska}, J.~X., \& {Davies}, R.~I.
  2022{\natexlab{b}}, \mnras, 509, 4457, \dodoi{10.1093/mnras/stab3232}

\bibitem[{{Belfiore} {et~al.}(2016){Belfiore}, {Maiolino}, {Maraston},
  {Emsellem}, {Bershady}, {Masters}, {Yan}, {Bizyaev}, {Boquien}, {Brownstein},
  {Bundy}, {Drory}, {Heckman}, {Law}, {Roman-Lopes}, {Pan}, {Stanghellini},
  {Thomas}, {Weijmans}, \& {Westfall}}]{Belfiore2016}
{Belfiore}, F., {Maiolino}, R., {Maraston}, C., {et~al.} 2016, \mnras, 461,
  3111, \dodoi{10.1093/mnras/stw1234}

\bibitem[{Belli {et~al.}(2019)Belli, Newman, \& Ellis}]{Belli2019}
Belli, S., Newman, A.~B., \& Ellis, R.~S. 2019, \apj, 874, 17,
  \dodoi{10.3847/1538-4357/ab07af}

\bibitem[{{Bellm} {et~al.}(2019){Bellm}, {Kulkarni}, {Graham}, {Dekany},
  {Smith}, {Riddle}, {Masci}, {Helou}, {Prince}, {Adams}, {Barbarino},
  {Barlow}, {Bauer}, {Beck}, {Belicki}, {Biswas}, {Blagorodnova}, {Bodewits},
  {Bolin}, {Brinnel}, {Brooke}, {Bue}, {Bulla}, {Burruss}, {Cenko}, {Chang},
  {Connolly}, {Coughlin}, {Cromer}, {Cunningham}, {De}, {Delacroix}, {Desai},
  {Duev}, {Eadie}, {Farnham}, {Feeney}, {Feindt}, {Flynn}, {Franckowiak},
  {Frederick}, {Fremling}, {Gal-Yam}, {Gezari}, {Giomi}, {Goldstein},
  {Golkhou}, {Goobar}, {Groom}, {Hacopians}, {Hale}, {Henning}, {Ho}, {Hover},
  {Howell}, {Hung}, {Huppenkothen}, {Imel}, {Ip}, {Ivezi{\'c}}, {Jackson},
  {Jones}, {Juric}, {Kasliwal}, {Kaspi}, {Kaye}, {Kelley}, {Kowalski},
  {Kramer}, {Kupfer}, {Landry}, {Laher}, {Lee}, {Lin}, {Lin}, {Lunnan},
  {Giomi}, {Mahabal}, {Mao}, {Miller}, {Monkewitz}, {Murphy}, {Ngeow},
  {Nordin}, {Nugent}, {Ofek}, {Patterson}, {Penprase}, {Porter}, {Rauch},
  {Rebbapragada}, {Reiley}, {Rigault}, {Rodriguez}, {van Roestel}, {Rusholme},
  {van Santen}, {Schulze}, {Shupe}, {Singer}, {Soumagnac}, {Stein}, {Surace},
  {Sollerman}, {Szkody}, {Taddia}, {Terek}, {Van Sistine}, {van Velzen},
  {Vestrand}, {Walters}, {Ward}, {Ye}, {Yu}, {Yan}, \& {Zolkower}}]{Bellm2019}
{Bellm}, E.~C., {Kulkarni}, S.~R., {Graham}, M.~J., {et~al.} 2019, \pasp, 131,
  018002, \dodoi{10.1088/1538-3873/aaecbe}

\bibitem[{{Bennert}(2005)}]{Bennert2005}
{Bennert}, V.~N. 2005, PhD thesis, Ruhr University, Bochum, Germany

\bibitem[{{Bezanson} {et~al.}(2022){Bezanson}, {Spilker}, {Suess}, {Setton},
  {Feldmann}, {Greene}, {Kriek}, {Narayanan}, \& {Verrico}}]{Bezanson2022}
{Bezanson}, R., {Spilker}, J.~S., {Suess}, K.~A., {et~al.} 2022, \apj, 925,
  153, \dodoi{10.3847/1538-4357/ac3dfa}

\bibitem[{{Brinchmann} {et~al.}(2004){Brinchmann}, {Charlot}, {White},
  {Tremonti}, {Kauffmann}, {Heckman}, \& {Brinkmann}}]{Brinchmann2004}
{Brinchmann}, J., {Charlot}, S., {White}, S.~D.~M., {et~al.} 2004, \mnras, 351,
  1151, \dodoi{10.1111/j.1365-2966.2004.07881.x}

\bibitem[{Brown {et~al.}(2009)Brown, Moustakas, Caldwell, Palamara, Cool, Dey,
  Hickox, Jannuzi, Murray, \& Zaritsky}]{Brown2009}
Brown, M. J.~I., Moustakas, J., Caldwell, N., {et~al.} 2009, \apj, 703, 150,
  \dodoi{10.1088/0004-637X/703/1/150}

\bibitem[{{Bundy} {et~al.}(2015){Bundy}, {Bershady}, {Law}, {Yan}, {Drory},
  {MacDonald}, {Wake}, {Cherinka}, {S{\'a}nchez-Gallego}, {Weijmans}, {Thomas},
  {Tremonti}, {Masters}, {Coccato}, {Diamond-Stanic}, {Arag{\'o}n-Salamanca},
  {Avila-Reese}, {Badenes}, {Falc{\'o}n-Barroso}, {Belfiore}, {Bizyaev},
  {Blanc}, {Bland-Hawthorn}, {Blanton}, {Brownstein}, {Byler}, {Cappellari},
  {Conroy}, {Dutton}, {Emsellem}, {Etherington}, {Frinchaboy}, {Fu}, {Gunn},
  {Harding}, {Johnston}, {Kauffmann}, {Kinemuchi}, {Klaene}, {Knapen},
  {Leauthaud}, {Li}, {Lin}, {Maiolino}, {Malanushenko}, {Malanushenko}, {Mao},
  {Maraston}, {McDermid}, {Merrifield}, {Nichol}, {Oravetz}, {Pan}, {Parejko},
  {Sanchez}, {Schlegel}, {Simmons}, {Steele}, {Steinmetz}, {Thanjavur},
  {Thompson}, {Tinker}, {van den Bosch}, {Westfall}, {Wilkinson}, {Wright},
  {Xiao}, \& {Zhang}}]{Bundy2015}
{Bundy}, K., {Bershady}, M.~A., {Law}, D.~R., {et~al.} 2015, \apj, 798, 7,
  \dodoi{10.1088/0004-637X/798/1/7}

\bibitem[{{Butler} \& {Bloom}(2011)}]{Butler2011}
{Butler}, N.~R., \& {Bloom}, J.~S. 2011, \aj, 141, 93,
  \dodoi{10.1088/0004-6256/141/3/93}

\bibitem[{Caldwell {et~al.}(1996)Caldwell, Rose, Franx, \&
  Leonardi}]{Caldwell1996}
Caldwell, N., Rose, J.~A., Franx, M., \& Leonardi, A.~J. 1996, The Astronomical
  Journal, 111, 78, \dodoi{10.1086/117762}

\bibitem[{Cales \& Brotherton(2015)}]{Cales2015}
Cales, S.~L., \& Brotherton, M.~S. 2015, \mnras, 449, 2374.
\newblock \doarXiv{1502.01086}

\bibitem[{{Chen} {et~al.}(2019){Chen}, {Shi}, {Dempsey}, {Law}, {Chen}, {Yan},
  {Bing}, {Rembold}, {Li}, {Yu}, {Riffel}, {Brownstein}, \&
  {Riffel}}]{Chen2019ENLR}
{Chen}, J., {Shi}, Y., {Dempsey}, R., {et~al.} 2019, \mnras, 489, 855,
  \dodoi{10.1093/mnras/stz2183}

\bibitem[{{Cherinka} {et~al.}(2019){Cherinka}, {Andrews},
  {S{\'a}nchez-Gallego}, {Brownstein}, {Argudo-Fern{\'a}ndez}, {Blanton},
  {Bundy}, {Jones}, {Masters}, {Law}, {Rowlands}, {Weijmans}, {Westfall}, \&
  {Yan}}]{Cherinka19}
{Cherinka}, B., {Andrews}, B.~H., {S{\'a}nchez-Gallego}, J., {et~al.} 2019,
  \aj, 158, 74, \dodoi{10.3847/1538-3881/ab2634}

\bibitem[{{Cid Fernandes} {et~al.}(2011){Cid Fernandes}, {Stasi{\'n}ska},
  {Mateus}, \& {Vale Asari}}]{CidFernandes2011}
{Cid Fernandes}, R., {Stasi{\'n}ska}, G., {Mateus}, A., \& {Vale Asari}, N.
  2011, \mnras, 413, 1687, \dodoi{10.1111/j.1365-2966.2011.18244.x}

\bibitem[{{Davies} {et~al.}(2022){Davies}, {Pontzen}, \& {Crain}}]{Davies2022}
{Davies}, J.~J., {Pontzen}, A., \& {Crain}, R.~A. 2022, \mnras, 515, 1430,
  \dodoi{10.1093/mnras/stac1742}

\bibitem[{Davies {et~al.}(2007)Davies, Sanchez, Genzel, Tacconi, Hicks,
  Friedrich, \& Sternberg}]{Davies2007}
Davies, R., Sanchez, F.~M., Genzel, R., {et~al.} 2007, \apj, 671, 1388,
  \dodoi{10.1086/523032}

\bibitem[{{De Propris} \& Melnick(2014)}]{DePropris2014}
{De Propris}, R., \& Melnick, J. 2014, \mnras, 439, 2837,
  \dodoi{10.1093/mnras/stu141}

\bibitem[{D'Eugenio {et~al.}(2020)D'Eugenio, van~der Wel, Wu, Barone, van
  Houdt, Bezanson, Straatman, Pacifici, Muzzin, Gallazzi, Wild, Sobral, Bell,
  Zibetti, Mowla, \& Franx}]{DEugenio2020}
D'Eugenio, F., van~der Wel, A., Wu, P.-F., {et~al.} 2020, MNRAS, 000, 1.
\newblock \doarXiv{2007.00663}

\bibitem[{{Dey} {et~al.}(2019){Dey}, {Schlegel}, {Lang}, {Blum}, {Burleigh},
  {Fan}, {Findlay}, {Finkbeiner}, {Herrera}, {Juneau}, {Landriau}, {Levi},
  {McGreer}, {Meisner}, {Myers}, {Moustakas}, {Nugent}, {Patej}, {Schlafly},
  {Walker}, {Valdes}, {Weaver}, {Y{\`e}che}, {Zou}, {Zhou}, {Abareshi},
  {Abbott}, {Abolfathi}, {Aguilera}, {Alam}, {Allen}, {Alvarez}, {Annis},
  {Ansarinejad}, {Aubert}, {Beechert}, {Bell}, {BenZvi}, {Beutler}, {Bielby},
  {Bolton}, {Brice{\~n}o}, {Buckley-Geer}, {Butler}, {Calamida}, {Carlberg},
  {Carter}, {Casas}, {Castander}, {Choi}, {Comparat}, {Cukanovaite}, {Delubac},
  {DeVries}, {Dey}, {Dhungana}, {Dickinson}, {Ding}, {Donaldson}, {Duan},
  {Duckworth}, {Eftekharzadeh}, {Eisenstein}, {Etourneau}, {Fagrelius},
  {Farihi}, {Fitzpatrick}, {Font-Ribera}, {Fulmer}, {G{\"a}nsicke},
  {Gaztanaga}, {George}, {Gerdes}, {Gontcho}, {Gorgoni}, {Green}, {Guy},
  {Harmer}, {Hernandez}, {Honscheid}, {Huang}, {James}, {Jannuzi}, {Jiang},
  {Joyce}, {Karcher}, {Karkar}, {Kehoe}, {Kneib}, {Kueter-Young}, {Lan},
  {Lauer}, {Le Guillou}, {Le Van Suu}, {Lee}, {Lesser}, {Perreault Levasseur},
  {Li}, {Mann}, {Marshall}, {Mart{\'\i}nez-V{\'a}zquez}, {Martini}, {du Mas des
  Bourboux}, {McManus}, {Meier}, {M{\'e}nard}, {Metcalfe},
  {Mu{\~n}oz-Guti{\'e}rrez}, {Najita}, {Napier}, {Narayan}, {Newman}, {Nie},
  {Nord}, {Norman}, {Olsen}, {Paat}, {Palanque-Delabrouille}, {Peng},
  {Poppett}, {Poremba}, {Prakash}, {Rabinowitz}, {Raichoor}, {Rezaie},
  {Robertson}, {Roe}, {Ross}, {Ross}, {Rudnick}, {Safonova}, {Saha},
  {S{\'a}nchez}, {Savary}, {Schweiker}, {Scott}, {Seo}, {Shan}, {Silva},
  {Slepian}, {Soto}, {Sprayberry}, {Staten}, {Stillman}, {Stupak}, {Summers},
  {Sien Tie}, {Tirado}, {Vargas-Maga{\~n}a}, {Vivas}, {Wechsler}, {Williams},
  {Yang}, {Yang}, {Yapici}, {Zaritsky}, {Zenteno}, {Zhang}, {Zhang}, {Zhou}, \&
  {Zhou}}]{Dey2019}
{Dey}, A., {Schlegel}, D.~J., {Lang}, D., {et~al.} 2019, \aj, 157, 168,
  \dodoi{10.3847/1538-3881/ab089d}

\bibitem[{{Ellison} {et~al.}(2022){Ellison}, {Wilkinson}, {Woo}, {Leung},
  {Wild}, {Bickley}, {Patton}, {Quai}, \& {Gwyn}}]{Ellison2022}
{Ellison}, S.~L., {Wilkinson}, S., {Woo}, J., {et~al.} 2022, \mnras, 517, L92,
  \dodoi{10.1093/mnrasl/slac109}

\bibitem[{{Elvis} {et~al.}(1994){Elvis}, {Wilkes}, {McDowell}, {Green},
  {Bechtold}, {Willner}, {Oey}, {Polomski}, \& {Cutri}}]{Elvis1994}
{Elvis}, M., {Wilkes}, B.~J., {McDowell}, J.~C., {et~al.} 1994, \apjs, 95, 1,
  \dodoi{10.1086/192093}

\bibitem[{{Evans} {et~al.}(2010){Evans}, {Primini}, {Glotfelty}, {Anderson},
  {Bonaventura}, {Chen}, {Davis}, {Doe}, {Evans}, {Fabbiano}, {Galle}, {Gibbs},
  {Grier}, {Hain}, {Hall}, {Harbo}, {He}, {Houck}, {Karovska}, {Kashyap},
  {Lauer}, {McCollough}, {McDowell}, {Miller}, {Mitschang}, {Morgan},
  {Mossman}, {Nichols}, {Nowak}, {Plummer}, {Refsdal}, {Rots}, {Siemiginowska},
  {Sundheim}, {Tibbetts}, {Van Stone}, {Winkelman}, \& {Zografou}}]{Evans2010}
{Evans}, I.~N., {Primini}, F.~A., {Glotfelty}, K.~J., {et~al.} 2010, \apjs,
  189, 37, \dodoi{10.1088/0067-0049/189/1/37}

\bibitem[{{Faber} {et~al.}(2007){Faber}, {Willmer}, {Wolf}, {Koo}, {Weiner},
  {Newman}, {Im}, {Coil}, {Conroy}, {Cooper}, {Davis}, {Finkbeiner}, {Gerke},
  {Gebhardt}, {Groth}, {Guhathakurta}, {Harker}, {Kaiser}, {Kassin},
  {Kleinheinrich}, {Konidaris}, {Kron}, {Lin}, {Luppino}, {Madgwick},
  {Meisenheimer}, {Noeske}, {Phillips}, {Sarajedini}, {Schiavon}, {Simard},
  {Szalay}, {Vogt}, \& {Yan}}]{Faber2007}
{Faber}, S.~M., {Willmer}, C.~N.~A., {Wolf}, C., {et~al.} 2007, \apj, 665, 265,
  \dodoi{10.1086/519294}

\bibitem[{{Florez} {et~al.}(2020){Florez}, {Jogee}, {Sherman}, {Stevans},
  {Finkelstein}, {Papovich}, {Kawinwanichakij}, {Ciardullo}, {Gronwall},
  {Urry}, {Kirkpatrick}, {LaMassa}, {Ananna}, \& {Wold}}]{Florez2020}
{Florez}, J., {Jogee}, S., {Sherman}, S., {et~al.} 2020, \mnras, 497, 3273,
  \dodoi{10.1093/mnras/staa2200}

\bibitem[{{French}(2021)}]{French2021}
{French}, K.~D. 2021, \pasp, 133, 072001, \dodoi{10.1088/1538-3873/ac0a59}

\bibitem[{{French} {et~al.}(2016){French}, {Arcavi}, \&
  {Zabludoff}}]{French2016}
{French}, K.~D., {Arcavi}, I., \& {Zabludoff}, A. 2016, \apjl, 818, L21,
  \dodoi{10.3847/2041-8205/818/1/L21}

\bibitem[{{French} {et~al.}(2020){French}, {Wevers}, {Law-Smith}, {Graur}, \&
  {Zabludoff}}]{French2020}
{French}, K.~D., {Wevers}, T., {Law-Smith}, J., {Graur}, O., \& {Zabludoff},
  A.~I. 2020, Space Science Reviews, 216, 32,
  \dodoi{10.1007/s11214-020-00657-y}

\bibitem[{French {et~al.}(2015)French, Yang, Zabludoff, Narayanan, Shirley,
  Walter, Smith, \& Tremonti}]{French2015}
French, K.~D., Yang, Y., Zabludoff, A., {et~al.} 2015, \apj, 801, 1,
  \dodoi{10.1088/0004-637X/801/1/1}

\bibitem[{French {et~al.}(2018)French, Yang, Zabludoff, \&
  Tremonti}]{French2018}
French, K.~D., Yang, Y., Zabludoff, A.~I., \& Tremonti, C.~A. 2018, \apj, 862,
  2, \dodoi{10.3847/1538-4357/AACB2D}

\bibitem[{{French} {et~al.}(2023){French}, {Smercina}, {Rowlands}, {Tripathi},
  {Zabludoff}, {Smith}, {Narayanan}, {Yang}, {Shirley}, \&
  {Alatalo}}]{French2023}
{French}, K.~D., {Smercina}, A., {Rowlands}, K., {et~al.} 2023, \apj, 942, 25,
  \dodoi{10.3847/1538-4357/aca46e}

\bibitem[{{Georgakakis} {et~al.}(2008){Georgakakis}, {Nandra}, {Yan},
  {Willner}, {Lotz}, {Pierce}, {Cooper}, {Laird}, {Koo}, {Barmby}, {Newman},
  {Primack}, \& {Coil}}]{Georgakakis2008}
{Georgakakis}, A., {Nandra}, K., {Yan}, R., {et~al.} 2008, \mnras, 385, 2049,
  \dodoi{10.1111/j.1365-2966.2008.12962.x}

\bibitem[{{Giustini} {et~al.}(2020){Giustini}, {Miniutti}, \&
  {Saxton}}]{Giustini2020}
{Giustini}, M., {Miniutti}, G., \& {Saxton}, R.~D. 2020, \aap, 636, L2,
  \dodoi{10.1051/0004-6361/202037610}

\bibitem[{{Goto}(2005)}]{Goto2005}
{Goto}, T. 2005, \mnras, 357, 937, \dodoi{10.1111/j.1365-2966.2005.08701.x}

\bibitem[{{Graur} {et~al.}(2018){Graur}, {French}, {Zahid}, {Guillochon},
  {Mandel}, {Auchettl}, \& {Zabludoff}}]{Graur2018}
{Graur}, O., {French}, K.~D., {Zahid}, H.~J., {et~al.} 2018, \apj, 853, 39,
  \dodoi{10.3847/1538-4357/aaa3fd}

\bibitem[{Greene {et~al.}(2020)Greene, Setton, Bezanson, Suess, Kriek, Spilker,
  Goulding, \& Feldmann}]{Greene2020}
Greene, J.~E., Setton, D., Bezanson, R., {et~al.} 2020, \apj, 899, L9,
  \dodoi{10.3847/2041-8213/aba534}

\bibitem[{{Greene} {et~al.}(2011){Greene}, {Zakamska}, {Ho}, \&
  {Barth}}]{Greene2011}
{Greene}, J.~E., {Zakamska}, N.~L., {Ho}, L.~C., \& {Barth}, A.~J. 2011, \apj,
  732, 9, \dodoi{10.1088/0004-637X/732/1/9}

\bibitem[{Harris {et~al.}(2020)Harris, Millman, van~der Walt, Gommers,
  Virtanen, Cournapeau, Wieser, Taylor, Berg, Smith, Kern, Picus, Hoyer, van
  Kerkwijk, Brett, Haldane, del R{\'{i}}o, Wiebe, Peterson,
  G{\'{e}}rard-Marchant, Sheppard, Reddy, Weckesser, Abbasi, Gohlke, \&
  Oliphant}]{numpy}
Harris, C.~R., Millman, K.~J., van~der Walt, S.~J., {et~al.} 2020, Nature, 585,
  357, \dodoi{10.1038/s41586-020-2649-2}

\bibitem[{{Harrison} {et~al.}(2014){Harrison}, {Alexander}, {Mullaney}, \&
  {Swinbank}}]{Harrison2014}
{Harrison}, C.~M., {Alexander}, D.~M., {Mullaney}, J.~R., \& {Swinbank}, A.~M.
  2014, \mnras, 441, 3306, \dodoi{10.1093/mnras/stu515}

\bibitem[{{Hickox} {et~al.}(2014){Hickox}, {Mullaney}, {Alexander}, {Chen},
  {Civano}, {Goulding}, \& {Hainline}}]{Hickox2014}
{Hickox}, R.~C., {Mullaney}, J.~R., {Alexander}, D.~M., {et~al.} 2014, \apj,
  782, 9, \dodoi{10.1088/0004-637X/782/1/9}

\bibitem[{{Hinkle} {et~al.}(2021){Hinkle}, {Holoien}, {Auchettl}, {Shappee},
  {Neustadt}, {Payne}, {Brown}, {Kochanek}, {Stanek}, {Graham}, {Tucker}, {Do},
  {Anderson}, {Bose}, {Chen}, {Coulter}, {Dimitriadis}, {Dong}, {Foley},
  {Huber}, {Hung}, {Kilpatrick}, {Pignata}, {Piro}, {Rojas-Bravo}, {Siebert},
  {Stalder}, {Thompson}, {Tonry}, {Vallely}, \& {Wisniewski}}]{Hinkle2021}
{Hinkle}, J.~T., {Holoien}, T.~W.~S., {Auchettl}, K., {et~al.} 2021, \mnras,
  500, 1673, \dodoi{10.1093/mnras/staa3170}

\bibitem[{{Ho} \& {Keto}(2007)}]{Ho2007}
{Ho}, L.~C., \& {Keto}, E. 2007, \apj, 658, 314, \dodoi{10.1086/511260}

\bibitem[{Hogg {et~al.}(2006)Hogg, Masjedi, Berlind, Blanton, Quintero, \&
  Brinkmann}]{Hogg2006}
Hogg, D.~W., Masjedi, M., Berlind, A.~A., {et~al.} 2006, \apj, 650, 763,
  \dodoi{10.1086/507172}

\bibitem[{Hopkins(2012)}]{Hopkins2012b}
Hopkins, P.~F. 2012, \mnras: Letters, 420, L8,
  \dodoi{10.1111/j.1745-3933.2011.01179.x}

\bibitem[{Hunter(2007)}]{matplotlib}
Hunter, J.~D. 2007, Computing in Science \& Engineering, 9, 90,
  \dodoi{10.1109/MCSE.2007.55}

\bibitem[{{Kauffmann} {et~al.}(2003{\natexlab{a}}){Kauffmann}, {Heckman},
  {Tremonti}, {Brinchmann}, {Charlot}, {White}, {Ridgway}, {Brinkmann},
  {Fukugita}, {Hall}, {Ivezi{\'c}}, {Richards}, \& {Schneider}}]{Kauffmann2003}
{Kauffmann}, G., {Heckman}, T.~M., {Tremonti}, C., {et~al.} 2003{\natexlab{a}},
  \mnras, 346, 1055, \dodoi{10.1111/j.1365-2966.2003.07154.x}

\bibitem[{{Kauffmann} {et~al.}(2003{\natexlab{b}}){Kauffmann}, {Heckman},
  {White}, {Charlot}, {Tremonti}, {Brinchmann}, {Bruzual}, {Peng}, {Seibert},
  {Bernardi}, {Blanton}, {Brinkmann}, {Castander}, {Cs{\'a}bai}, {Fukugita},
  {Ivezic}, {Munn}, {Nichol}, {Padmanabhan}, {Thakar}, {Weinberg}, \&
  {York}}]{Kauffmann2003a}
{Kauffmann}, G., {Heckman}, T.~M., {White}, S. D.~M., {et~al.}
  2003{\natexlab{b}}, \mnras, 341, 33, \dodoi{10.1046/j.1365-8711.2003.06291.x}

\bibitem[{{Keel} {et~al.}(2012){Keel}, {Chojnowski}, {Bennert}, {Schawinski},
  {Lintott}, {Lynn}, {Pancoast}, {Harris}, {Nierenberg}, {Sonnenfeld}, \&
  {Proctor}}]{Keel2012}
{Keel}, W.~C., {Chojnowski}, S.~D., {Bennert}, V.~N., {et~al.} 2012, \mnras,
  420, 878, \dodoi{10.1111/j.1365-2966.2011.20101.x}

\bibitem[{{Keel} {et~al.}(2017){Keel}, {Lintott}, {Maksym}, {Bennert},
  {Chojnowski}, {Moiseev}, {Smirnova}, {Schawinski}, {Sartori}, {Urry},
  {Pancoast}, {Schirmer}, {Scott}, {Showley}, \& {Flatland}}]{Keel2017}
{Keel}, W.~C., {Lintott}, C.~J., {Maksym}, W.~P., {et~al.} 2017, \apj, 835,
  256, \dodoi{10.3847/1538-4357/835/2/256}

\bibitem[{{Kennedy} {et~al.}(2016){Kennedy}, {Meiron}, {Shukirgaliyev},
  {Panamarev}, {Berczik}, {Just}, \& {Spurzem}}]{Kennedy2016}
{Kennedy}, G.~F., {Meiron}, Y., {Shukirgaliyev}, B., {et~al.} 2016, \mnras,
  460, 240, \dodoi{10.1093/mnras/stw908}

\bibitem[{{Kewley} {et~al.}(2001){Kewley}, {Dopita}, {Sutherland}, {Heisler},
  \& {Trevena}}]{Kewley2001}
{Kewley}, L.~J., {Dopita}, M.~A., {Sutherland}, R.~S., {Heisler}, C.~A., \&
  {Trevena}, J. 2001, \apj, 556, 121, \dodoi{10.1086/321545}

\bibitem[{{Kewley} {et~al.}(2006){Kewley}, {Groves}, {Kauffmann}, \&
  {Heckman}}]{Kewley2006}
{Kewley}, L.~J., {Groves}, B., {Kauffmann}, G., \& {Heckman}, T. 2006, \mnras,
  372, 961, \dodoi{10.1111/j.1365-2966.2006.10859.x}

\bibitem[{{King}(2020)}]{King2020}
{King}, A. 2020, \mnras, 493, L120, \dodoi{10.1093/mnrasl/slaa020}

\bibitem[{{Komossa} \& {Schulz}(1997)}]{Komossa1997}
{Komossa}, S., \& {Schulz}, H. 1997, \aap, 323, 31,
  \dodoi{10.48550/arXiv.astro-ph/9701001}

\bibitem[{{Lanz} {et~al.}(2022){Lanz}, {Stepanoff}, {Hickox}, {Alatalo},
  {French}, {Rowlands}, {Nyland}, {Appleton}, {Lacy}, {Medling}, {Mulchaey},
  {Sazonova}, \& {Urry}}]{Lanz2022}
{Lanz}, L., {Stepanoff}, S., {Hickox}, R.~C., {et~al.} 2022, \apj, 935, 29,
  \dodoi{10.3847/1538-4357/ac7d56}

\bibitem[{Li {et~al.}(2019)Li, French, Zabludoff, \& Ho}]{Li2019}
Li, Z., French, K.~D., Zabludoff, A.~I., \& Ho, L.~C. 2019, \apj, 879, 131,
  \dodoi{10.3847/1538-4357/ab1f68}

\bibitem[{{Lintott} {et~al.}(2009){Lintott}, {Schawinski}, {Keel}, {van Arkel},
  {Bennert}, {Edmondson}, {Thomas}, {Smith}, {Herbert}, {Jarvis}, {Virani},
  {Andreescu}, {Bamford}, {Land}, {Murray}, {Nichol}, {Raddick}, {Slosar},
  {Szalay}, \& {Vandenberg}}]{Lintott2009}
{Lintott}, C.~J., {Schawinski}, K., {Keel}, W., {et~al.} 2009, \mnras, 399,
  129, \dodoi{10.1111/j.1365-2966.2009.15299.x}

\bibitem[{{Liu} {et~al.}(2013){Liu}, {Zakamska}, {Greene}, {Nesvadba}, \&
  {Liu}}]{Liu2013}
{Liu}, G., {Zakamska}, N.~L., {Greene}, J.~E., {Nesvadba}, N. P.~H., \& {Liu},
  X. 2013, \mnras, 430, 2327, \dodoi{10.1093/mnras/stt051}

\bibitem[{{Lotz} {et~al.}(2021){Lotz}, {Dolag}, {Remus}, \&
  {Burkert}}]{Lotz2021}
{Lotz}, M., {Dolag}, K., {Remus}, R.-S., \& {Burkert}, A. 2021, \mnras, 506,
  4516, \dodoi{10.1093/mnras/stab2037}

\bibitem[{{Luridiana} {et~al.}(2015){Luridiana}, {Morisset}, \&
  {Shaw}}]{Luridiana2015}
{Luridiana}, V., {Morisset}, C., \& {Shaw}, R.~A. 2015, \aap, 573, A42,
  \dodoi{10.1051/0004-6361/201323152}

\bibitem[{{Marconi} {et~al.}(2004){Marconi}, {Risaliti}, {Gilli}, {Hunt},
  {Maiolino}, \& {Salvati}}]{Marconi2004}
{Marconi}, A., {Risaliti}, G., {Gilli}, R., {et~al.} 2004, \mnras, 351, 169,
  \dodoi{10.1111/j.1365-2966.2004.07765.x}

\bibitem[{{McElroy} {et~al.}(2015){McElroy}, {Croom}, {Pracy}, {Sharp}, {Ho},
  \& {Medling}}]{McElroy2015}
{McElroy}, R., {Croom}, S.~M., {Pracy}, M., {et~al.} 2015, \mnras, 446, 2186,
  \dodoi{10.1093/mnras/stu2224}

\bibitem[{Mendel {et~al.}(2013)Mendel, Simard, Ellison, \& Patton}]{Mendel2013}
Mendel, J.~T., Simard, L., Ellison, S.~L., \& Patton, D.~R. 2013, \mnras, 429,
  2212, \dodoi{10.1093/mnras/sts489}

\bibitem[{{Metzger} {et~al.}(2022){Metzger}, {Stone}, \&
  {Gilbaum}}]{Metzger2022}
{Metzger}, B.~D., {Stone}, N.~C., \& {Gilbaum}, S. 2022, \apj, 926, 101,
  \dodoi{10.3847/1538-4357/ac3ee1}

\bibitem[{{Meusinger} {et~al.}(2017){Meusinger}, {Br{\"u}necke}, {Schalldach},
  \& {in der Au}}]{Meusinger2016}
{Meusinger}, H., {Br{\"u}necke}, J., {Schalldach}, P., \& {in der Au}, A. 2017,
  \aap, 597, A134, \dodoi{10.1051/0004-6361/201629139}

\bibitem[{{Neugebauer} {et~al.}(1984){Neugebauer}, {Habing}, {van Duinen},
  {Aumann}, {Baud}, {Beichman}, {Beintema}, {Boggess}, {Clegg}, {de Jong},
  {Emerson}, {Gautier}, {Gillett}, {Harris}, {Hauser}, {Houck}, {Jennings},
  {Low}, {Marsden}, {Miley}, {Olnon}, {Pottasch}, {Raimond}, {Rowan-Robinson},
  {Soifer}, {Walker}, {Wesselius}, \& {Young}}]{Neugebauer1984}
{Neugebauer}, G., {Habing}, H.~J., {van Duinen}, R., {et~al.} 1984, \apjl, 278,
  L1, \dodoi{10.1086/184209}

\bibitem[{{Pawlik} {et~al.}(2016){Pawlik}, {Wild}, {Walcher}, {Johansson},
  {Villforth}, {Rowlands}, {Mendez-Abreu}, \& {Hewlett}}]{Pawlik2016}
{Pawlik}, M.~M., {Wild}, V., {Walcher}, C.~J., {et~al.} 2016, \mnras, 456,
  3032, \dodoi{10.1093/mnras/stv2878}

\bibitem[{{Pawlik} {et~al.}(2018){Pawlik}, {Taj Aldeen}, {Wild},
  {Mendez-Abreu}, {Lah{\'e}n}, {Johansson}, {Jimenez}, {Lucas}, {Zheng},
  {Walcher}, \& {Rowlands}}]{Pawlik2018}
{Pawlik}, M.~M., {Taj Aldeen}, L., {Wild}, V., {et~al.} 2018, \mnras, 477,
  1708, \dodoi{10.1093/mnras/sty589}

\bibitem[{{Pontzen} {et~al.}(2017){Pontzen}, {Tremmel}, {Roth}, {Peiris},
  {Saintonge}, {Volonteri}, {Quinn}, \& {Governato}}]{Pontzen2017}
{Pontzen}, A., {Tremmel}, M., {Roth}, N., {et~al.} 2017, \mnras, 465, 547,
  \dodoi{10.1093/mnras/stw2627}

\bibitem[{Pracy {et~al.}(2009)Pracy, Kuntschner, Couch, Blake, Bekki, \&
  Briggs}]{Pracy2009}
Pracy, M.~B., Kuntschner, H., Couch, W.~J., {et~al.} 2009, \mnras, 396, 1349,
  \dodoi{10.1111/j.1365-2966.2009.14836.x}

\bibitem[{{Prieto} {et~al.}(2016){Prieto}, {Kr{\"u}hler}, {Anderson},
  {Galbany}, {Kochanek}, {Aquino}, {Brown}, {Dong}, {F{\"o}rster}, {Holoien},
  {Kuncarayakti}, {Maureira}, {Rosales-Ortega}, {S{\'a}nchez}, {Shappee}, \&
  {Stanek}}]{Prieto2016}
{Prieto}, J.~L., {Kr{\"u}hler}, T., {Anderson}, J.~P., {et~al.} 2016, \apjl,
  830, L32, \dodoi{10.3847/2041-8205/830/2/L32}

\bibitem[{Rich {et~al.}(2015)Rich, Kewley, \& Dopita}]{Rich2015}
Rich, J.~A., Kewley, L.~J., \& Dopita, M.~A. 2015, \apjs, 221, 28,
  \dodoi{10.1088/0067-0049/221/2/28}

\bibitem[{Rowlands {et~al.}(2015)Rowlands, Wild, Nesvadba, Sibthorpe, Mortier,
  Lehnert, \& da~Cunha}]{Rowlands2015}
Rowlands, K., Wild, V., Nesvadba, N., {et~al.} 2015, \mnras, 448, 258,
  \dodoi{10.1093/mnras/stu2714}

\bibitem[{{Saintonge} {et~al.}(2011){Saintonge}, {Kauffmann}, {Kramer},
  {Tacconi}, {Buchbender}, {Catinella}, {Fabello}, {Graci{\'a}-Carpio}, {Wang},
  {Cortese}, {Fu}, {Genzel}, {Giovanelli}, {Guo}, {Haynes}, {Heckman},
  {Krumholz}, {Lemonias}, {Li}, {Moran}, {Rodriguez-Fernandez}, {Schiminovich},
  {Schuster}, \& {Sievers}}]{Saintonge2011}
{Saintonge}, A., {Kauffmann}, G., {Kramer}, C., {et~al.} 2011, \mnras, 415, 32,
  \dodoi{10.1111/j.1365-2966.2011.18677.x}

\bibitem[{{Salim} {et~al.}(2007){Salim}, {Rich}, {Charlot}, {Brinchmann},
  {Johnson}, {Schiminovich}, {Seibert}, {Mallery}, {Heckman}, {Forster},
  {Friedman}, {Martin}, {Morrissey}, {Neff}, {Small}, {Wyder}, {Bianchi},
  {Donas}, {Lee}, {Madore}, {Milliard}, {Szalay}, {Welsh}, \& {Yi}}]{Salim2007}
{Salim}, S., {Rich}, R.~M., {Charlot}, S., {et~al.} 2007, \apjs, 173, 267,
  \dodoi{10.1086/519218}

\bibitem[{{S{\'a}nchez} {et~al.}(2018){S{\'a}nchez}, {Avila-Reese},
  {Hernandez-Toledo}, {Cortes-Su{\'a}rez}, {Rodr{\'\i}guez-Puebla},
  {Ibarra-Medel}, {Cano-D{\'\i}az}, {Barrera-Ballesteros}, {Negrete},
  {Calette}, {de Lorenzo-C{\'a}ceres}, {Ortega-Minakata}, {Aquino},
  {Valenzuela}, {Clemente}, {Storchi-Bergmann}, {Riffel}, {Schimoia}, {Riffel},
  {Rembold}, {Brownstein}, {Pan}, {Yates}, {Mallmann}, \&
  {Bitsakis}}]{Sanchez2018}
{S{\'a}nchez}, S.~F., {Avila-Reese}, V., {Hernandez-Toledo}, H., {et~al.} 2018,
  \rmxaa, 54, 217, \dodoi{10.48550/arXiv.1709.05438}

\bibitem[{{Sartori} {et~al.}(2018{\natexlab{a}}){Sartori}, {Schawinski},
  {Trakhtenbrot}, {Caplar}, {Treister}, {Koss}, {Urry}, \&
  {Zhang}}]{Sartori2018}
{Sartori}, L.~F., {Schawinski}, K., {Trakhtenbrot}, B., {et~al.}
  2018{\natexlab{a}}, \mnras, 476, L34, \dodoi{10.1093/mnrasl/sly025}

\bibitem[{{Sartori} {et~al.}(2018{\natexlab{b}}){Sartori}, {Schawinski},
  {Koss}, {Ricci}, {Treister}, {Stern}, {Lansbury}, {Maksym}, {Balokovi{\'c}},
  {Gandhi}, {Keel}, \& {Ballantyne}}]{Sartori2018b}
{Sartori}, L.~F., {Schawinski}, K., {Koss}, M.~J., {et~al.} 2018{\natexlab{b}},
  \mnras, 474, 2444, \dodoi{10.1093/mnras/stx2952}

\bibitem[{{Sazonova} {et~al.}(2021){Sazonova}, {Alatalo}, {Rowlands},
  {Deustua}, {French}, {Heckman}, {Lanz}, {Lisenfeld}, {Luo}, {Medling},
  {Nyland}, {Otter}, {Petric}, {Snyder}, \& {Urry}}]{Sazonova2021}
{Sazonova}, E., {Alatalo}, K., {Rowlands}, K., {et~al.} 2021, arXiv e-prints,
  arXiv:2105.09956.
\newblock \doarXiv{2105.09956}

\bibitem[{Schawinski {et~al.}(2009)Schawinski, Virani, Simmons, Urry, Treister,
  Kaviraj, \& Kushkuley}]{Schawinski2009}
Schawinski, K., Virani, S., Simmons, B., {et~al.} 2009, \apj, 692, L19,
  \dodoi{10.1088/0004-637X/692/1/L19}

\bibitem[{{Schawinski} {et~al.}(2010){Schawinski}, {Urry}, {Virani}, {Coppi},
  {Bamford}, {Treister}, {Lintott}, {Sarzi}, {Keel}, {Kaviraj}, {Cardamone},
  {Masters}, {Ross}, {Andreescu}, {Murray}, {Nichol}, {Raddick}, {Slosar},
  {Szalay}, {Thomas}, \& {Vandenberg}}]{Schawinski2010}
{Schawinski}, K., {Urry}, C.~M., {Virani}, S., {et~al.} 2010, \apj, 711, 284,
  \dodoi{10.1088/0004-637X/711/1/284}

\bibitem[{Schweizer {et~al.}(2013)Schweizer, Seitzer, Kelson, Villanueva, \&
  Walth}]{Schweizer2013}
Schweizer, F., Seitzer, P., Kelson, D.~D., Villanueva, E.~V., \& Walth, G.~L.
  2013, \apj, 773, 148, \dodoi{10.1088/0004-637X/773/2/148}

\bibitem[{{Shankar} {et~al.}(2009){Shankar}, {Weinberg}, \&
  {Miralda-Escud{\'e}}}]{Shankar2009}
{Shankar}, F., {Weinberg}, D.~H., \& {Miralda-Escud{\'e}}, J. 2009, \apj, 690,
  20, \dodoi{10.1088/0004-637X/690/1/20}

\bibitem[{{Silk} \& {Rees}(1998)}]{Silk1998}
{Silk}, J., \& {Rees}, M.~J. 1998, \aap, 331, L1.
\newblock \doarXiv{astro-ph/9801013}

\bibitem[{Smercina {et~al.}(2018)Smercina, Smith, Dale, French, Croxall,
  Zhukovska, Togi, Bell, Crocker, Draine, Jarrett, Tremonti, Yang, \&
  Zabludoff}]{Smercina2018}
Smercina, A., Smith, J. D.~T., Dale, D.~A., {et~al.} 2018, \apj, 855, 51,
  \dodoi{10.3847/1538-4357/aaafcd}

\bibitem[{{Smercina} {et~al.}(2022){Smercina}, {Smith}, {French}, {Bell},
  {Dale}, {Medling}, {Nyland}, {Privon}, {Rowlands}, {Walter}, \&
  {Zabludoff}}]{Smercina2022}
{Smercina}, A., {Smith}, J.-D.~T., {French}, K.~D., {et~al.} 2022, \apj, 929,
  154, \dodoi{10.3847/1538-4357/ac5d5f}

\bibitem[{Snyder {et~al.}(2011)Snyder, Cox, Hayward, Hernquist, \&
  Jonsson}]{Snyder2011}
Snyder, G.~F., Cox, T.~J., Hayward, C.~C., Hernquist, L., \& Jonsson, P. 2011,
  \apj, 741, 77, \dodoi{10.1088/0004-637X/741/2/77}

\bibitem[{{Springel} {et~al.}(2005){Springel}, {Di Matteo}, \&
  {Hernquist}}]{Springel2005}
{Springel}, V., {Di Matteo}, T., \& {Hernquist}, L. 2005, \apjl, 620, L79,
  \dodoi{10.1086/428772}

\bibitem[{{Stern} {et~al.}(2012){Stern}, {Assef}, {Benford}, {Blain}, {Cutri},
  {Dey}, {Eisenhardt}, {Griffith}, {Jarrett}, {Lake}, {Masci}, {Petty},
  {Stanford}, {Tsai}, {Wright}, {Yan}, {Harrison}, \& {Madsen}}]{Stern2012}
{Stern}, D., {Assef}, R.~J., {Benford}, D.~J., {et~al.} 2012, \apj, 753, 30,
  \dodoi{10.1088/0004-637X/753/1/30}

\bibitem[{{Strateva} {et~al.}(2001){Strateva}, {Ivezi{\'c}}, {Knapp},
  {Narayanan}, {Strauss}, {Gunn}, {Lupton}, {Schlegel}, {Bahcall}, {Brinkmann},
  {Brunner}, {Budav{\'a}ri}, {Csabai}, {Castander}, {Doi}, {Fukugita},
  {Gy{\H{o}}ry}, {Hamabe}, {Hennessy}, {Ichikawa}, {Kunszt}, {Lamb}, {McKay},
  {Okamura}, {Racusin}, {Sekiguchi}, {Schneider}, {Shimasaku}, \&
  {York}}]{Strateva2001}
{Strateva}, I., {Ivezi{\'c}}, {\v{Z}}., {Knapp}, G.~R., {et~al.} 2001, \aj,
  122, 1861, \dodoi{10.1086/323301}

\bibitem[{{Strauss} {et~al.}(2002){Strauss}, {Weinberg}, {Lupton}, {Narayanan},
  {Annis}, {Bernardi}, {Blanton}, {Burles}, {Connolly}, {Dalcanton}, {Doi},
  {Eisenstein}, {Frieman}, {Fukugita}, {Gunn}, {Ivezi{\'c}}, {Kent}, {Kim},
  {Knapp}, {Kron}, {Munn}, {Newberg}, {Nichol}, {Okamura}, {Quinn}, {Richmond},
  {Schlegel}, {Shimasaku}, {SubbaRao}, {Szalay}, {Vanden Berk}, {Vogeley},
  {Yanny}, {Yasuda}, {York}, \& {Zehavi}}]{Strauss2002}
{Strauss}, M.~A., {Weinberg}, D.~H., {Lupton}, R.~H., {et~al.} 2002, \aj, 124,
  1810, \dodoi{10.1086/342343}

\bibitem[{Suess {et~al.}(2017)Suess, Bezanson, Spilker, Kriek, Greene,
  Feldmann, Hunt, \& Narayanan}]{Suess2017}
Suess, K.~A., Bezanson, R., Spilker, J.~S., {et~al.} 2017, \apjl, 846, L14,
  \dodoi{10.3847/2041-8213/aa85dc}

\bibitem[{{Toft} {et~al.}(2014){Toft}, {Smol{\v{c}}i{\'c}}, {Magnelli},
  {Karim}, {Zirm}, {Michalowski}, {Capak}, {Sheth}, {Schawinski}, {Krogager},
  {Wuyts}, {Sanders}, {Man}, {Lutz}, {Staguhn}, {Berta}, {Mccracken}, {Krpan},
  \& {Riechers}}]{Toft2014}
{Toft}, S., {Smol{\v{c}}i{\'c}}, V., {Magnelli}, B., {et~al.} 2014, \apj, 782,
  68, \dodoi{10.1088/0004-637X/782/2/68}

\bibitem[{{Tremonti} {et~al.}(2004){Tremonti}, {Heckman}, {Kauffmann},
  {Brinchmann}, {Charlot}, {White}, {Seibert}, {Peng}, {Schlegel}, {Uomoto},
  {Fukugita}, \& {Brinkmann}}]{Tremonti2004}
{Tremonti}, C.~A., {Heckman}, T.~M., {Kauffmann}, G., {et~al.} 2004, \apj, 613,
  898, \dodoi{10.1086/423264}

\bibitem[{{Veilleux} \& {Osterbrock}(1987)}]{veilleux87}
{Veilleux}, S., \& {Osterbrock}, D.~E. 1987, \apjs, 63, 295,
  \dodoi{10.1086/191166}

\bibitem[{{Wang} {et~al.}(2022){Wang}, {Yin}, {Ma}, \& {Wu}}]{Wang2022}
{Wang}, M., {Yin}, J., {Ma}, Y., \& {Wu}, Q. 2022, \apj, 933, 225,
  \dodoi{10.3847/1538-4357/ac75e6}

\bibitem[{{Watkins} {et~al.}(2018){Watkins}, {Mihos}, {Bershady}, \&
  {Harding}}]{Watkins2018}
{Watkins}, A.~E., {Mihos}, J.~C., {Bershady}, M., \& {Harding}, P. 2018, \apjl,
  858, L16, \dodoi{10.3847/2041-8213/aabba1}

\bibitem[{{Westfall} {et~al.}(2019){Westfall}, {Cappellari}, {Bershady},
  {Bundy}, {Belfiore}, {Ji}, {Law}, {Schaefer}, {Shetty}, {Tremonti}, {Yan},
  {Andrews}, {Brownstein}, {Cherinka}, {Coccato}, {Drory}, {Maraston},
  {Parikh}, {S{\'a}nchez-Gallego}, {Thomas}, {Weijmans}, {Barrera-Ballesteros},
  {Du}, {Goddard}, {Li}, {Masters}, {Ibarra Medel}, {S{\'a}nchez}, {Yang},
  {Zheng}, \& {Zhou}}]{Westfall2019}
{Westfall}, K.~B., {Cappellari}, M., {Bershady}, M.~A., {et~al.} 2019, \aj,
  158, 231, \dodoi{10.3847/1538-3881/ab44a2}

\bibitem[{{Wevers} {et~al.}(2022){Wevers}, {Pasham}, {Jalan}, {Rakshit}, \&
  {Arcodia}}]{Wevers2022}
{Wevers}, T., {Pasham}, D.~R., {Jalan}, P., {Rakshit}, S., \& {Arcodia}, R.
  2022, \aap, 659, L2, \dodoi{10.1051/0004-6361/202243143}

\bibitem[{Wild {et~al.}(2010)Wild, Heckman, \& Charlot}]{Wild2010}
Wild, V., Heckman, T., \& Charlot, S. 2010, \mnras, 405, 933,
  \dodoi{10.1111/j.1365-2966.2010.16536.x}

\bibitem[{Wild {et~al.}(2007)Wild, Kauffmann, Heckman, Charlot, Lemson,
  Brinchmann, Reichard, \& Pasquali}]{Wild2007}
Wild, V., Kauffmann, G., Heckman, T., {et~al.} 2007, \mnras, 381, 543,
  \dodoi{10.1111/j.1365-2966.2007.12256.x}

\bibitem[{Wild {et~al.}(2009)Wild, Walcher, Johansson, Tresse, Charlot, Pollo,
  {Le F{\`{e}}vre}, \& de~Ravel}]{Wild2009}
Wild, V., Walcher, C.~J., Johansson, P.~H., {et~al.} 2009, \mnras, 395, 144,
  \dodoi{10.1111/j.1365-2966.2009.14537.x}

\bibitem[{{Wild} {et~al.}(2020){Wild}, {Taj Aldeen}, {Carnall}, {Maltby},
  {Almaini}, {Werle}, {Wilkinson}, {Rowlands}, {Bolzonella}, {Castellano},
  {Gargiulo}, {McLure}, {Pentericci}, \& {Pozzetti}}]{Wild2020}
{Wild}, V., {Taj Aldeen}, L., {Carnall}, A., {et~al.} 2020, \mnras, 494, 529,
  \dodoi{10.1093/mnras/staa674}

\bibitem[{{Wilkinson} {et~al.}(2021){Wilkinson}, {Almaini}, {Wild}, {Maltby},
  {Hartley}, {Simpson}, \& {Rowlands}}]{Wilkinson2021}
{Wilkinson}, A., {Almaini}, O., {Wild}, V., {et~al.} 2021, \mnras,
  \dodoi{10.1093/mnras/stab965}

\bibitem[{{Wuyts} {et~al.}(2011){Wuyts}, {F{\"o}rster Schreiber}, {van der
  Wel}, {Magnelli}, {Guo}, {Genzel}, {Lutz}, {Aussel}, {Barro}, {Berta},
  {Cava}, {Graci{\'a}-Carpio}, {Hathi}, {Huang}, {Kocevski}, {Koekemoer},
  {Lee}, {Le Floc'h}, {McGrath}, {Nordon}, {Popesso}, {Pozzi}, {Riguccini},
  {Rodighiero}, {Saintonge}, \& {Tacconi}}]{Wuyts2011}
{Wuyts}, S., {F{\"o}rster Schreiber}, N.~M., {van der Wel}, A., {et~al.} 2011,
  \apj, 742, 96, \dodoi{10.1088/0004-637X/742/2/96}

\bibitem[{{Xu} \& {Wang}(2023)}]{Xu2023}
{Xu}, X., \& {Wang}, J. 2023, \apj, 943, 28, \dodoi{10.3847/1538-4357/acac82}

\bibitem[{Yan \& Blanton(2012)}]{Yan2012}
Yan, R., \& Blanton, M.~R. 2012, \apj, 747, 61,
  \dodoi{10.1088/0004-637X/747/1/61}

\bibitem[{Yan {et~al.}(2006)Yan, Newman, Faber, Konidaris, Koo, \&
  Davis}]{Yan2006}
Yan, R., Newman, J.~A., Faber, S.~M., {et~al.} 2006, \apj, 648, 281,
  \dodoi{10.1086/505629}

\bibitem[{{Yan} {et~al.}(2016){Yan}, {Bundy}, {Law}, {Bershady}, {Andrews},
  {Cherinka}, {Diamond-Stanic}, {Drory}, {MacDonald}, {S{\'a}nchez-Gallego},
  {Thomas}, {Wake}, {Weijmans}, {Westfall}, {Zhang}, {Arag{\'o}n-Salamanca},
  {Belfiore}, {Bizyaev}, {Blanc}, {Blanton}, {Brownstein}, {Cappellari},
  {D'Souza}, {Emsellem}, {Fu}, {Gaulme}, {Graham}, {Goddard}, {Gunn},
  {Harding}, {Jones}, {Kinemuchi}, {Li}, {Li}, {Maiolino}, {Mao}, {Maraston},
  {Masters}, {Merrifield}, {Oravetz}, {Pan}, {Parejko}, {Sanchez}, {Schlegel},
  {Simmons}, {Thanjavur}, {Tinker}, {Tremonti}, {van den Bosch}, \&
  {Zheng}}]{Yan2016}
{Yan}, R., {Bundy}, K., {Law}, D.~R., {et~al.} 2016, \aj, 152, 197,
  \dodoi{10.3847/0004-6256/152/6/197}

\bibitem[{Yang {et~al.}(2006)Yang, Tremonti, Zabludoff, \& Zaritsky}]{Yang2006}
Yang, Y., Tremonti, C.~A., Zabludoff, A.~I., \& Zaritsky, D. 2006, \apj, 646,
  L33, \dodoi{10.1086/506909}

\bibitem[{Yang {et~al.}(2004)Yang, Zabludoff, Zaritsky, Lauer, \&
  Mihos}]{Yang2004}
Yang, Y., Zabludoff, A.~I., Zaritsky, D., Lauer, T.~R., \& Mihos, J.~C. 2004,
  \apj, 607, 258, \dodoi{10.1086/383259}

\bibitem[{Yang {et~al.}(2008)Yang, Zabludoff, Zaritsky, \& Mihos}]{Yang2008}
Yang, Y., Zabludoff, A.~I., Zaritsky, D., \& Mihos, J.~C. 2008, \apj, 688, 945,
  \dodoi{10.1086/591656}

\bibitem[{{York} {et~al.}(2000){York}, {Adelman}, {Anderson}, {Anderson},
  {Annis}, {Bahcall}, {Bakken}, {Barkhouser}, {Bastian}, {Berman}, {Boroski},
  {Bracker}, {Briegel}, {Briggs}, {Brinkmann}, {Brunner}, {Burles}, {Carey},
  {Carr}, {Castander}, {Chen}, {Colestock}, {Connolly}, {Crocker}, {Csabai},
  {Czarapata}, {Davis}, {Doi}, {Dombeck}, {Eisenstein}, {Ellman}, {Elms},
  {Evans}, {Fan}, {Federwitz}, {Fiscelli}, {Friedman}, {Frieman}, {Fukugita},
  {Gillespie}, {Gunn}, {Gurbani}, {de Haas}, {Haldeman}, {Harris}, {Hayes},
  {Heckman}, {Hennessy}, {Hindsley}, {Holm}, {Holmgren}, {Huang}, {Hull},
  {Husby}, {Ichikawa}, {Ichikawa}, {Ivezi{\'c}}, {Kent}, {Kim}, {Kinney},
  {Klaene}, {Kleinman}, {Kleinman}, {Knapp}, {Korienek}, {Kron}, {Kunszt},
  {Lamb}, {Lee}, {Leger}, {Limmongkol}, {Lindenmeyer}, {Long}, {Loomis},
  {Loveday}, {Lucinio}, {Lupton}, {MacKinnon}, {Mannery}, {Mantsch}, {Margon},
  {McGehee}, {McKay}, {Meiksin}, {Merelli}, {Monet}, {Munn}, {Narayanan},
  {Nash}, {Neilsen}, {Neswold}, {Newberg}, {Nichol}, {Nicinski}, {Nonino},
  {Okada}, {Okamura}, {Ostriker}, {Owen}, {Pauls}, {Peoples}, {Peterson},
  {Petravick}, {Pier}, {Pope}, {Pordes}, {Prosapio}, {Rechenmacher}, {Quinn},
  {Richards}, {Richmond}, {Rivetta}, {Rockosi}, {Ruthmansdorfer}, {Sandford},
  {Schlegel}, {Schneider}, {Sekiguchi}, {Sergey}, {Shimasaku}, {Siegmund},
  {Smee}, {Smith}, {Snedden}, {Stone}, {Stoughton}, {Strauss}, {Stubbs},
  {SubbaRao}, {Szalay}, {Szapudi}, {Szokoly}, {Thakar}, {Tremonti}, {Tucker},
  {Uomoto}, {Vanden Berk}, {Vogeley}, {Waddell}, {Wang}, {Watanabe},
  {Weinberg}, {Yanny}, {Yasuda}, \& {SDSS Collaboration}}]{York2000}
{York}, D.~G., {Adelman}, J., {Anderson}, John~E., J., {et~al.} 2000, \aj, 120,
  1579, \dodoi{10.1086/301513}

\bibitem[{Zabludoff {et~al.}(1996)Zabludoff, Zaritsky, Lin, Tucker, Hashimoto,
  Shectman, Oemler, \& Kirshner}]{Zabludoff1996}
Zabludoff, A.~I., Zaritsky, D., Lin, H., {et~al.} 1996, \apj, 466, 104,
  \dodoi{10.1086/177495}

\bibitem[{{Zolotov} {et~al.}(2015){Zolotov}, {Dekel}, {Mandelker}, {Tweed},
  {Inoue}, {DeGraf}, {Ceverino}, {Primack}, {Barro}, \& {Faber}}]{Zolotov2015}
{Zolotov}, A., {Dekel}, A., {Mandelker}, N., {et~al.} 2015, \mnras, 450, 2327,
  \dodoi{10.1093/mnras/stv740}

\end{thebibliography}

\end{document}